\documentclass[preprint,aps,prl,showpacs,preprintnumbers,amsmath,amssymb,superscriptaddress]{revtex4-1}
\usepackage{graphicx}  
\usepackage{bm}        
\usepackage{amssymb,multirow}   
\usepackage[usenames,dvipsnames]{xcolor}
\usepackage[hyperfootnotes=false]{hyperref}
\usepackage[hyperfootnotes=false]{hyperref}
\hypersetup{  colorlinks=true,  citecolor=Violet,  linkcolor=Blue,  filecolor= green,  urlcolor=Blue}
\usepackage{dcolumn}
\usepackage{bm}
\usepackage{multirow}
\usepackage{subfigure}
\usepackage{xcolor}
\usepackage[normalem]{ulem}
\usepackage{soul}

\begin{document}

\author{Shishir Kumar Pandey}
\thanks{Contributed equally to this work}
\author{Abhinav Kumar}
\thanks{Contributed equally to this work}
\author{Sagar Sarkar}
\author{Priya Mahadevan}
\affiliation{Department of Condensed Matter Physics and Material Science, \\ S. N. Bose National Center for Basic Sciences,Kolkata}
\email{priya.mahadevan@gmail.com}

\title{Understanding the ferromagnetic insulating state in Cr doped VO$_2$}

\begin{abstract}
Experimentally Cr doping in the rutile phase of VO$_2$ is found to stabilize a charge ordered 
ferromagnetic insulating state 
in the doping range of 10\% to 20\%. In this work, we investigated its origin  at 12.5\% Cr doping  using a combination of 
ab-initio electronic structure calculations as well as microscopic modeling. Our calculations are found to 
reproduce the ferromagnetic insulating state as well as a charge ordering at the V and Cr sites. The 
mapping of the ab-initio band structure onto a tight-binding Hamiltonian 
allows one to calculate the energy
gain from different exchange pathways. This gain is quantified in this work for the first time and
the role of charge ordering in stabilizing a ferromagnetic insulating state is understood.
\end{abstract}

\maketitle

\section{Introduction}
Among the magnetic systems that one encounters, an empirical rule has emerged, which is that ferromagnetism is accompanied by a metallic ground state while antiferromagnetism is found 
in members that are insulating. The ones that break these empirical trends are the most interesting members as they would require a mechanism beyond the conventionally accepted 
theories to explain the origin of the magnetic state. 
Even from a technological standpoint ferromagnetic insulators are interesting   \cite{fmi_appl1,fmi_appl2} as information transfer can take place through spin waves, without 
charge displacement or eddy current losses. There have been several examples discovered 
recently which include the layered materials like CrI$_3$, VI$_3$ \cite{cri3,vi3} as well as a three-dimensional oxide
 Sr$_3$OsO$_6$ with a high ordering temperature \cite{sroso}. The recently discovered
layered van der Waals ferromagnet Cr$_2$Ge$_2$Te$_6$ is also found to be insulating \cite{crgete}. 
However, the mechanism here as in several other compounds such as CdCr$_2$S$_4$  is 90$^o$ superexchange between the Cr atoms via the S atom \cite{cdcrs}. 
This allows a ferromagnetic exchange pathway even in insulators. Ferromagnetic insulators 
are also found among the double perovskites of the form A$_2$BB$'$O$_6$. Here, the presence of different atoms at the B and B$'$ sites, where the B$'$ site is usually nonmagnetic, leads to a superexchange pathway through the unoccupied states on the B' atom that is operational even in an
insulator. Even when the B$'$ site is magnetic, we have few examples of ferromagnetic insulators as found in Bi$_2$NiMnO$_6$, La$_2$NiMnO$_6$ for
instance \cite{bnmo,lnmo}. The highly distorted perovskite oxide YTiO$_3$ was found to have a ferromagnetic ground state because of orbital ordering at the Ti
site, which also explained the insulating character of the ground state \cite{ytio}. There have been other examples found among the undoped oxides such
as strained films of LaCoO$_3$, though the mechanism has not been clarified there\cite{lcoo}.  
The doped manganites \cite{manganites} as well as the hollandites K$_2$Cr$_8$O$_{16}$ have an unusual ground state where charge ordering is found to co-exist with a ferromagnetic
insulating state \cite{kcro_expt}. In Ref. \cite{kcro_th} it was speculated that the charge ordering led to the presence of a superexchange pathway that could stabilize 
a ferromagnetic insulator. The present study allows us to examine the role played by charge ordering quantitatively.

While V$_{1-x}$Cr$_x$O$_2$ \cite{mono-26,mono-16,poupet6} in its monoclinic phase has been  extensively studied
for its metal-insulator transitions, the more recent stabilization of the rutile phase in Cr doped 
VO$_2$ \cite{exp,piper} led to the observation of a ferromagnetic insulating state. 
In this work, we examine the origin of this unusual ground state.
The parent compounds VO$_2$ \cite{vo2_exp1,tc6,mot6,peierl6} and CrO$_2$ \cite{cro2_exp1,cro2_exp2,cro2_exp3,cro2_exp4} are examples of an antiferromagnetic insulator (though in a different polymorph) and ferromagnetic metal respectively. 
However, experiments have found that for Cr doping percentage from 10\% to 20\% 
in the rutile phase of VO$_2$, the system is 
found to become a ferromagnetic insulator \cite{piper}. X-ray absorption spectroscopy which, apart from being atom specific is also sensitive to the valence state, reveals the presence of Cr$^{+3}$-V$^{+5}$ pairs \cite{piper}. 
The question that follows is why do such charge states get stabilized and do they have a role
in the unusual ferromagnetic insulating state found at this doping concentration. 

V atoms have a valence state of +4 in VO$_2$, with the $d$ electrons on V 
having an electronic configuration of $d^1$. An isovalent substitution of 
V with Cr should result in a configuration of $d^2$ at the Cr sites. As the
fully doped end-member CrO$_2$ is metallic, one expects a metallic ground 
state to be favoured above a critical doping concentration. While the
transport properties of these systems haven't been studied, x-ray
absorption spectroscopy results which are sensitive to the valence state
of the transition metal atoms reveal the presence of V$^{+5}$ as well as
Cr$^{+3}$ species, in addition to V$^{+4}$ at 18 \% doping of Cr  \cite{piper}. 
These results are suggestive of an insulating ground 
state being favoured, as otherwise the free carriers present would result
in all the V atoms having the same oxidation state. 

In this work we have doped Cr into the experimentally observed rutile phase of VO$_2$ for doping
concentrations of 12.5 and 25 \% and examined the ensuing electronic and magnetic ground states. 
A ferromagnetic insulating state is found at 12.5\% which is consistent with the dopant range in experiments
where it has been seen. The doped Cr atoms are found to be in the +3 valence state instead of +4
which is expected for an isovalent substitution. This is achieved by the transfer of an electron 
from one of the neighboring V sites, rendering the latter with a valency of +5. While the
distortion of the Cr-O and V-O bondlengths  involve a large component of strain energy to 
stabilize the unusual valencies of Cr$^{+3}$ and V$^{+5}$, the large Hund's 
intraatomic exchange on Cr favouring a $d^3$ configuration as well as the attractive Coulomb 
interactions between Cr$^{+3}$-V$^{+5}$ ions help in stabilising it. The charge ordering of Cr$^{+3}$-V$^{+5}$ ions 
also facilitates hopping if the spins are aligned ferromagnetically
and is remniscent of a 'frozen-in' double exchange configuration. We carry out 
a mapping of the ab-initio band structure onto a tight-binding model for the ferromagnetic case as well as the
closest lying antiferromagnetic state at 12.5 \% doping.  
Within the model we determine the energy gain from
the Cr$^{+3}$-V$^{+5}$ pathway as well as Cr$^{+3}$-V$^{+4}$ pathway and show for the first time how charge ordering can stabilize
a ferromagnetic insulating state. This mechanism has to compete with other pathways present. 

At 25\% doping, the complexity of the possible competing magnetic ground states increases. This is because while the V atoms would prefer an antiferromagnetic ordering among themselves, the relative spin alignment of spins between Cr and V atoms depends on the separation between the Cr 
atoms. The half-filled $t_{2g}$ bands at the Cr$^{3+}$ site would like to be antiferromagnetic, while the pathway through V$^{5+}$ 
would favour a ferromagnetic ground state. Examining the lowest energy configuration here, we find it to be again a ferromagnetic insulator.

\section{Methodology}
In order to calculate the electronic structure of Cr doped VO$_2$, we have performed first principle density functional theory based calculations using Vienna $ab$ $initio$ simulation
package \cite{vasp}. We have used projected augmented wave \cite{paw} potentials for each atom. These include the semi-core $p$ states on the V atom. We have used a gamma-centered k-mesh   
of $6\times6\times8$ for the 24 atom unit cell considered, which has been appropriately scaled for the larger supercell sizes considered.
A plane wave cutoff of 875 eV was used in the calculations. The generalized gradient approximation (GGA)\cite{GGA} was used for the
exchange-correlation functional and electron-electron interactions were considered by including a Hubbard $U$ within the GGA+U formalism\cite{Dudarev}. One finds a variation of the $U$ from 2 to 4 eV for V and Cr in the literature \cite{uvaryref1,uvaryref2,uvaryref3}. This 
prompted us to explore variations of $U$ in this range, the details of which are given later in the manuscript. Having explored 
that the results are robust against variations in $U$, the rest of the discussion in the manuscript uses 
a $U$ of 2.5 eV on V atoms and a $U$ of 3 eV on the Cr atoms.  

The total energy was calculated self-consistently till the energy difference between successive steps was better than 10$^{-5}$ eV. The total energy of the ferromagnetic configuration was compared with the energy of other antiferromagnetic configurations within the 8-V atom $\sqrt{2}{\bf a}\times\sqrt{2}{\bf b}\times2${\bf c} supercell constructed using the experimental lattice parameters  of VO$_2$ as the starting parameters, where {\bf a} $=$ {\bf b} $=$ 6.441 \AA{} and {\bf c} $=$ 5.700 \AA{}  \cite{mcwhan}. We 
further optimized both the lattice parameters as well as the internal coordinates of the crystal. After optimization we found that {\bf a } and {\bf b} were slightly  ( approximately 0.6 - 0.7 \%) underestimated while {\bf c} was overestimated by 4.4 \% from it's initial value (Please see Table~S1 of supplementary information~\cite{footnote-SI} for more details). As these results correspond to the specific case where we have 
Cr$^{+3}$-V$^{+5}$ pairs
arranged alternately in the $c$-direction, we explored a larger unit cell  
$\sqrt{2}{\bf a}\times\sqrt{2}{\bf b}\times4{\bf c}$ in which we doped pairs of Cr atoms. All distinct  
locations for the pair of Cr atoms were explored.

In order to understand the results further we have mapped  the $ab$ $initio$ 
band structure  onto a tight binding model using maximally localized wannier functions for the radial part of the wavefunctions 
using WANNIER90-VASP interface \cite{wannier1,wannier2} implemented within VASP. 
The mapping was done for the 12.5 \% doping concentration which had 1 Cr atom in the unit cell and was done for both the
ferromagnetic as well as the closest lying antiferromagnetic state. 
The basis we considered for the tight binding model included all the five $d$ 
orbitals on Cr and V atoms as well as the three $p$ orbitals on oxygen atoms. A  sufficiently large energy window spanning from -7.8 eV below E$_f$ to 4.2 eV above it was chosen to accommodate all
the 88 Wannier functions of our basis. The spreads of each of the wannier functions are given in bottom of Table~S1 of supplementary information~\cite{footnote-SI}, with the largest being found to be 0.547, 0.611 and 
0.716 \AA{}$^2$ for Cr and V - $d$ and O-$p$ orbitals respectively. Consequently one can say that the wannier functions are sufficiently localized and representative of the
atomic-like orbitals.
The mapping then allows us to extract the onsite energies 
as well as the hopping interaction strengths. Our band structure in the wannier basis provides a very good description of the ab-initio band
structure. In Fig. S1 \cite{footnote-SI} we show the comparison when we retain interactions upto just 3.4 $\AA$ and see that retaining interactions till here provides us with a reasonable description of the electronic structure.
As we now have the complete Hamiltonian, we 
switch off certain hopping pathways and determine the band energy, computed as the sum of the occupied eigenvalues, 
within the tight binding model. This allows us to determine the contribution from various exchange pathways in each magnetic configuration. 

\section{Results and Discussion}

\begin{figure}[ht]
\centering
\includegraphics[height=7.0cm,width=12.0cm]{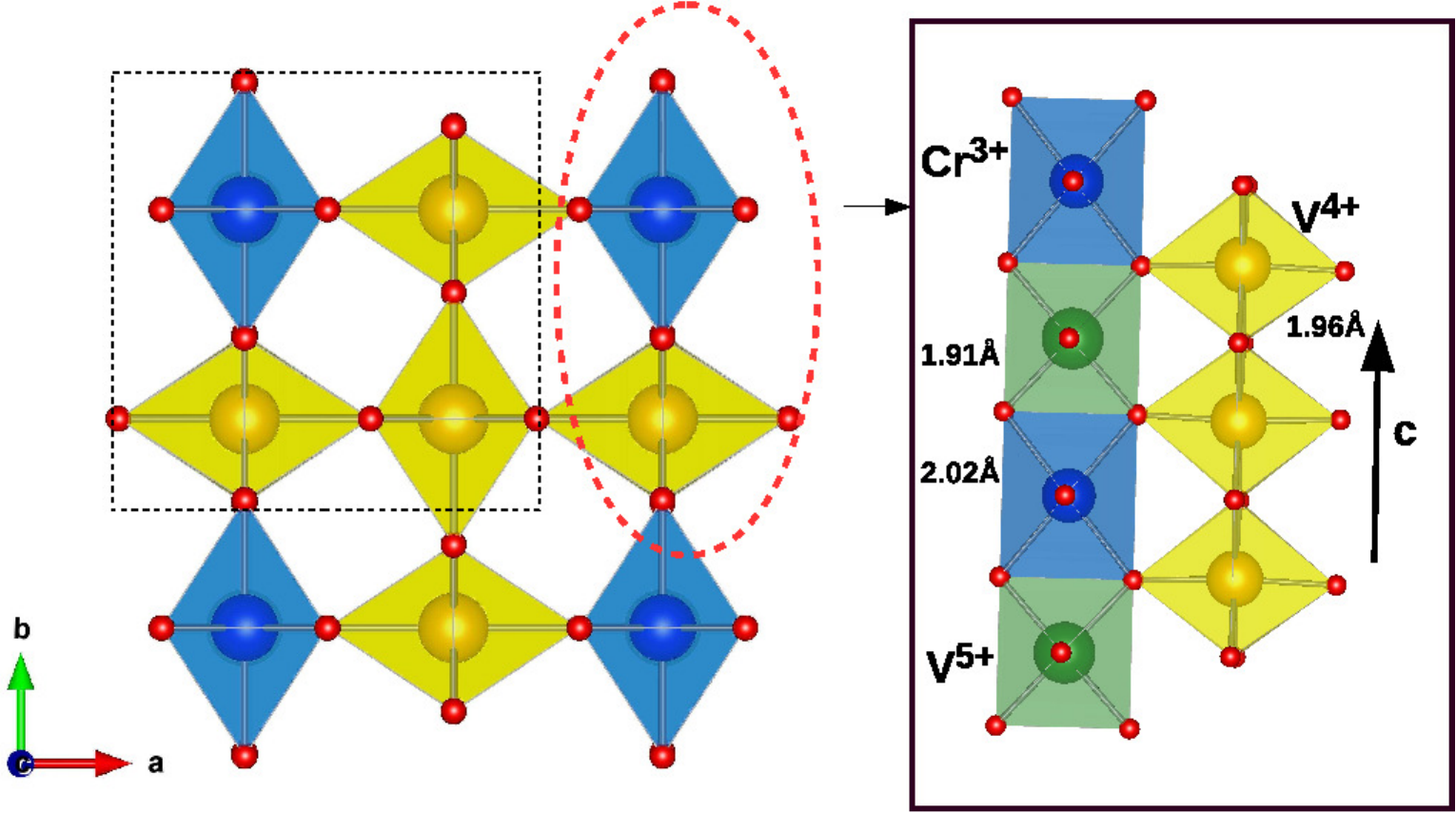}
\caption{The $\sqrt{2} {\bf a} \times  \sqrt{2} {\bf a} \times  2{\bf c}$ supercell has been shown on the  left while the formation of Cr$^{+3}$-V$^{+5}$ chains as well as the bond lengths have been shown in 
the right for 12.5 \% Cr doped in VO$_2$ (24 atom supercell). The blue and green balls represent Cr$^{+3}$ and V$^{+5}$ while V$^{+4}$ ions are represented by the 
yellow balls. Dashed black line square box show the rutile unit cell of VO$_2$ used to make the supercell considered in our study.  Dashed red oval box show arrangement of 
Cr$^{+3}$ -V$^{+4}$ chains along $b$-direction. }
  \label{fig.1}
\end{figure} 

\begin{figure}[ht]
\centering
 \includegraphics[height=5.0cm,width=12.0cm]{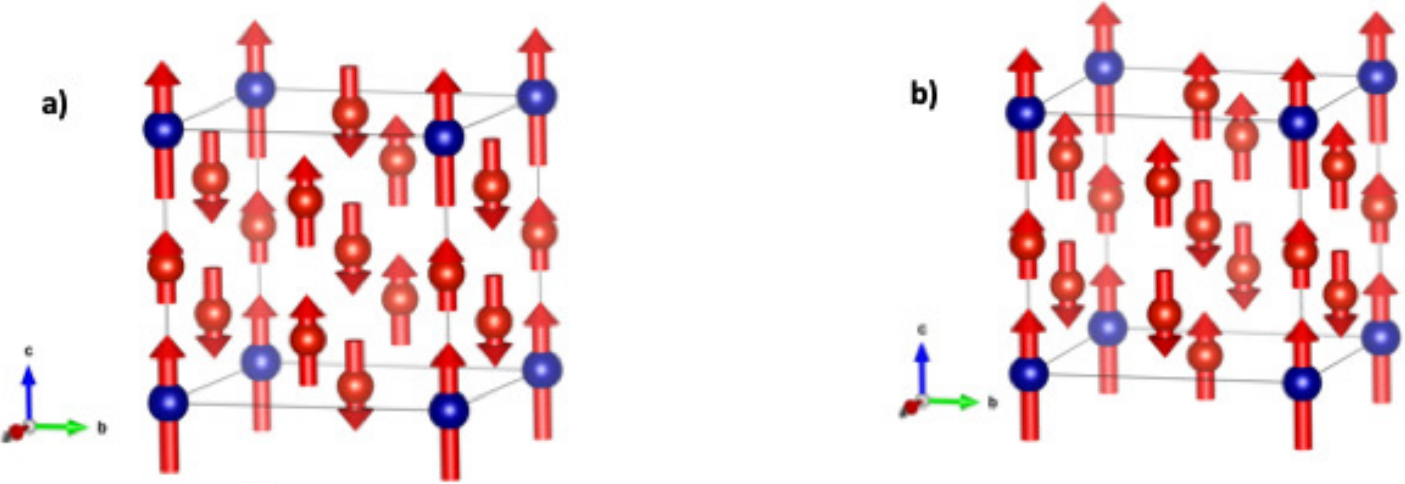}
 \caption{(a) AFM1 and (b) AFM2 magnetic configurations for 12.5 \% Cr doping case. Blue balls are Cr atoms and red ones are V. Length of the arrow
represents the corresponding magnitude of magnetic moments. Small red
arrow correspond to V$^{+5}$ ions with a tiny magnetic moment of 0.06 $\mu_B$. }
  \label{fig.2}
\end{figure}

In order to understand the electronic and magnetic properties of 
V$_{1-x}$Cr$_{x}$O$_2$, we consider a supercell of VO$_2$ which has eight V atoms in it,
shown in the left panel of Fig.~\ref{fig.1}. 
Each unit cell has four V chains running in the 
$c$-direction. This is also the direction in which the distance between
neighboring V atoms is the shortest and is equal to 2.94 \AA{} in the undoped case.
The other neighbouring V-V bondlengths are equal to
3.52 \AA{}. In order to explore the modifications in the electronic structure at 12.5\% doping, 
one V atom in this unit cell was
replaced by a Cr atom. Displacing the atoms and minimising their total energy, we found that the V-V distances are now modified 
and become 2.98 \AA{} along the c-direction, while the other  V-V bondlengths are 3.52/3.53 \AA{}.
In order to find  the ground state magnetic configuration, 
we examined the ferromagnetic as well as the antiferromagnetic configurations shown in Fig.~\ref{fig.2}.

In the antiferromagnetic configuration labelled AFM1, the coupling between the atoms in a chain is ferromagnetic. The neighbouring chains
are however coupled antiferromagnetically. 
The other antiferromagnetic configuration that was explored had the V atoms
coupled antiferromagnetically along the chain, while they were ferromagnetic between nearest neighbor chains. This was labelled AFM2.
The unit cell parameters were optimised for each of the magnetic configurations 
considered. They did not show very large changes between the different 
configurations. The total energies determined for the ferromagnetic
as well as various antiferromagnetic configurations are listed in Table~\ref{ch_CrVO2_tab.1}. 
The ferromagnetic configuration is found to be more stable than the closest 
antiferromagnetic configuration AFM1 by 100 meV per supercell.
As we have a doped Cr atom which carries a different moment than the V atoms, we find a net moment 
for each of the antiferromagnetic configurations probed and this has
been indicated in Table ~\ref{ch_CrVO2_tab.1}.  While the 
experimental ground state has not been examined at this composition, 
it has been found to be a ferromagnetic insulator at 10, 18  and 20 \%.
The presence of a ferromagnetic insulating state seems unusual, so we examine the
density of states at 12.5$\%$ where we already found a ferromagnetic 
ground state to be stabilized to see if one finds an insulating state also. 
\begin{table}[ht]
  \centering

  \begin{tabular}{c c c}
   \hline \hline
     & Energy (eV) & MM ($\mu_B$)\\
   \hline
    FM & 0.0  & 9 \\
    AFM1 & 0.1 & 1.0 \\
    AFM2 & 0.12 & 3.0 \\
    \hline \hline
  \end{tabular}
    \caption{Relative energies of different antiferromagnetic magnetic configurations per supercell (24 atom/8 formula units) with respect to ferromagnetic configuration for 12.5 \% doping of Cr in rutile VO$_2$. 
    Net magnetic moment (MM) is also given for each case.} 

    \label{ch_CrVO2_tab.1}
\end{table}

\begin{figure}[h!]
\centering
\vspace{20pt}
\includegraphics[height=7.0cm,width=10.0cm]{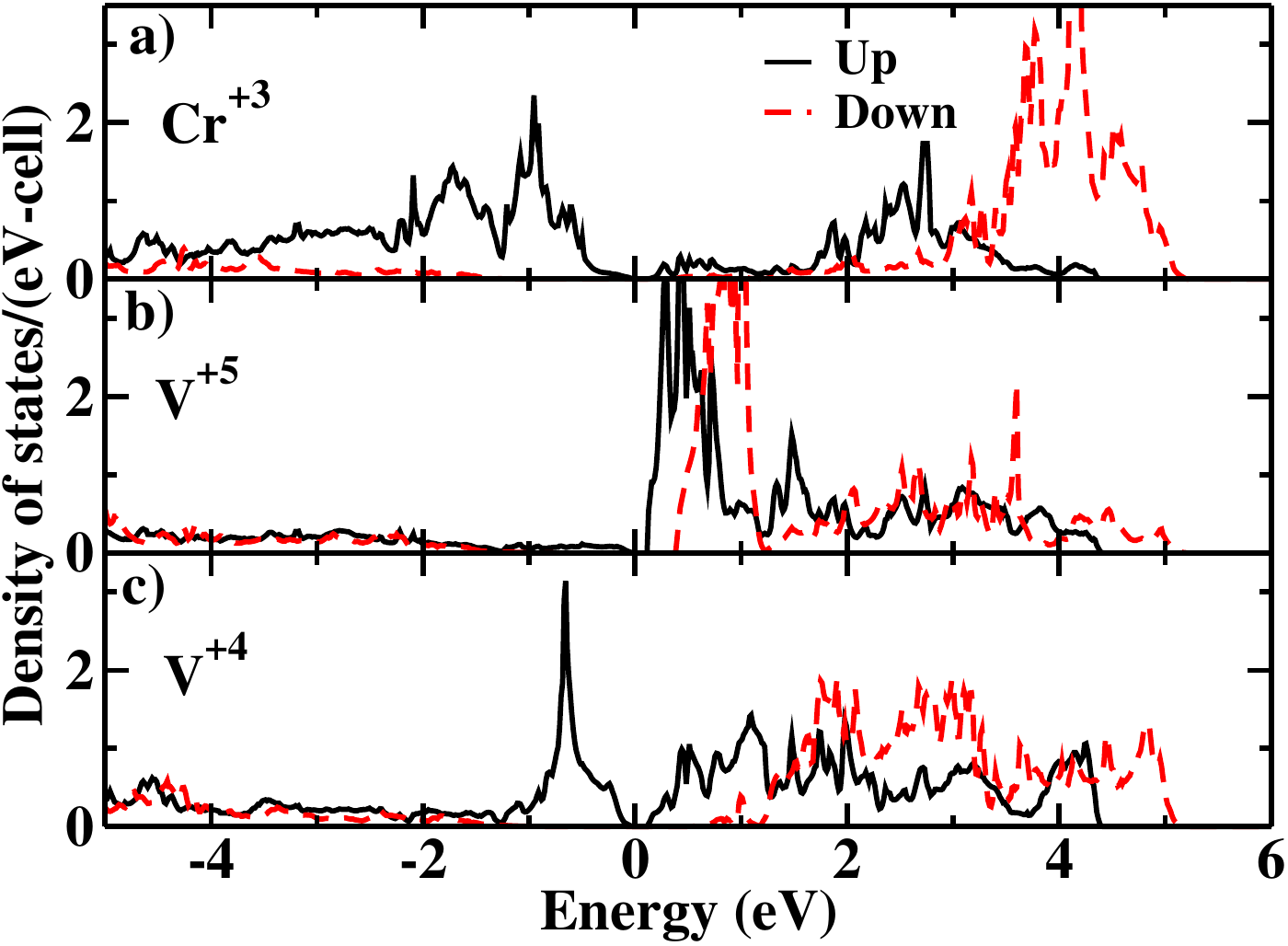}
\caption{Calculated spin projected transition metal $d$ density of states  for ferromagnetic Cr doped in rutile VO$_2$ at 12.5\% doping (24 atom supercell). Different panels correspond to the atoms identified as (a) Cr$^{+3}$, (b) V$^{+5}$ and 
(c) V$^{+4}$ in the calculations.}
  \label{fig.3}
\end{figure}

The atom, angular momentum and spin projected density of states are 
shown for two distinct V atoms that we find in the unit cell as well 
as the Cr atom. The system is found to be insulating. 
As each transition metal atom is surrounded by six oxygens, the $d$ orbitals at each transition 
metal site split into triply degenerate $t_{2g}$ orbitals at lower energies and doubly 
degenerate $e_g$ orbitals at higher energies. Examining the  Cr $d$ density of 
states shown in Fig.~\ref{fig.3}, one finds that the minority spin
states lie beyond 3~eV above the fermi level and are empty. The majority spin $t_{2g}$ states are however
found to be occupied while the $e_g$ counterparts are found to be empty. An isovalent 
substitution of V by Cr would imply a valence state of +4 on Cr. However, we find the electron 
configuration of $t_{2g\uparrow}^3$ on Cr, implying a valence of +3. One of the V atoms has 
none of the  $d$ orbitals occupied and hence can be associated with a $d^0$
configuration. We identify a valence of +5 with this V atom. Examining the density of states
associated with the other V atoms, we find that one can associate a valence of +4 with them.

The transition metal - oxygen bondlengths better reflect the oxidation state when we use bond-valence concepts.
The Cr-O bond lengths in the optimized structure were found to be 1.96 \AA{} ($\times$ 2)  and 2.02 \AA{} ($\times$ 4).
The presence of the longer bondlengths reflect the fact that the Cr atom has more electrons, and so minimises the Coulomb repulsion between
the electrons on Cr and those on O.
All the V-O bond lengths were 1.96 \AA{} in undoped VO$_2$. 
The V$^{+4}$-O bond lengths remain more or less similar to the parent compound with two bonds a little 
shorter (1.92/1.93 \AA{}). As one would expect shorter 
V-O bond lengths for a higher valency of V atom, we also found the V$^{+5}$-O bond lengths to be 1.90 \AA{} ($\times$ 6), smaller than that of the V$^{+4}$-O ones. The  V$^{4+}$ and V$^{5+}$ states correspond to a  formal $d^1$ and $d^0$ configuration at the V site. As the number of electrons is higher at the V site for the V$^{4+}$ site, here, the V-O oxygen bonds have to be longer than between V$^{5+}$ atoms and oxygen. This leads to larger hopping
interaction strengths in the latter case as against the former. Additionally the electrons in the higher valent state feel a stronger attraction from the nuclear potential. This leads to the $d$ levels being stabilised more, and consequently a smaller charge transfer energy for the higher valent V atom. Both these
factors contribute to stronger interactions between the V and the oxygen sites. Hence in this strongly hybridized picture, one ends up with
very small differences in the number of electrons between the different valent states. With a small charge transfer energy as found here, one
could also have a higher electron count for V$^{5+}$ than for the V$^{4+}$ atom.  \cite{oxd_state}.
The integrated transition metal $d$ component of the charge within spheres of radii 1.33 \AA{} for Cr and 1.323 \AA{} for V atoms 
on the sites labeled V$^{+4}$, V$^{+5}$ and Cr$^{+3}$ are found to be 
3.35, 3.38 and 4.16. While a formal electron
count of 1, 0 and 3 are expected in the transition metal $d$ orbitals based on the oxidation state, one finds the charge to
deviate significantly from these values. The difficulty in partitioning charge in charge
ordered systems has been discussed in the literature \cite{co_prl,oxd_state,oxd_state2,oxd_state3} 
and we repeat the discussion for completion.
The magnetic moments on the other hand are found to be 1.15, 0.21, 2.95 $\mu_B$.  They reflect the
valence state better as has been seen earlier \cite{co_prl}.  
However, here we find that the moment on V$^{+4}$ is higher than the value of
1 expected for the $d^1$ configuration that we have at the V site. This happens because the bare charge transfer energy is small and is equal
to 0.5 eV and becomes negative when we include the bandwidth of the O $p$ states. This leads to the enhanced magnetic moment for V$^{+4}$ and
V$^{+5}$ ions, over what is expected from a formal $d^1$ and $d^0$ configuration at the V sites. 
\begin{table}[ht]
\setlength{\tabcolsep}{2pt}
  \centering
 \begin{tabular}{c c c c c c c c r}
   \hline \hline
     & & $U$ (Cr, V) = 4, 3 eV  & & & & &  $U$ (Cr, V) = 5, 4 eV  & \\
   \cline{2-4} \cline{6-9}
     & Energy (eV)  & Mom ($\mu_B$) & Gap (eV)  & & &  Energy (eV)  & Mom ($\mu_B$) & Gap (eV)    \\
   \hline
    FM & 0.0  & 9.0 & 0.377 & &  & 0.0 & 9.0 & 0.795  \\
    AFM1 & 0.081 & 1.0 & 0.413 &  && 0.064 & 1.0 & 0.876 \\
    \hline \hline
  \end{tabular}
    \caption{Comparison of stability of ferromagnetic ground state and
the next competing ground state AFM1 at 12.5 \% doping in the 24 atom supercell with variation
of $U$ on Cr/V-$d$ states. The relative energies as well as the magnetic moment (Mom) and band gap 
have been given.}
    \label{tab.2}
\end{table}

The presence of charge ordering as well as the emergence of a ferromagnetic insulating state seems interesting, so we went on to examine if these results were a consequence of the choice of $U$ on the V and Cr sites. As the literature \cite {uvaryref1,uvaryref2,uvaryref3} reports a range of values, we varied $U$
(U$_{Cr}$/U$_V$) from 3/2.5 eV to 5/4 eV. In every case we find the existence of V$^{+4}$, V$^{+5}$ and Cr$^{+3}$ ions. In order to examine the stability of the ferromagnetic insulating state, we again compute the energies for different magnetic ground states.  Relative stability  of the FM-I ground state compared to the competing antiferromagnetic AFM1 configuration  as a function of $U$ 
are listed in Table~\ref{tab.2}. In every case we find that the ferromagnetic state is stabilized and the conclusions remain unchanged. Hybrid functional calculations were also performed where the component of the Hartree-Fock exchange was chosen to be 25\% as is usually considered for transition metal oxides. Here also we found a ferromagnetic insulating state to 
be stabilised by 0.094 eV over the closest lying antiferromagnetic state. The corresponding
density of states plot are given in Fig.~S2 of supplementary information~\cite{footnote-SI}. 
Magnetic moments on Cr$^{+3}$ and V$^{+4}$/V$^{+5}$ were found to be 2.91 and 1.12/0.16 $\mu_B$
respectively. The band gap was found to be 1.122 eV in contrast to the value of 0.1 eV we had for a $U$ of 3 eV on Cr and 2.5 eV on V, suggesting a larger value of $U$. We have not tried to determine the value of $U$ that would give us this band gap, as there is no consensus in the literature \cite{hse-alpha1,hse-alpha2} on the appropriate value of the Hartree-Fock exchange for VO$_2$. 
\begin{table}[ht]
\setlength{\tabcolsep}{2pt}
  \centering
  \begin{tabular}{c c c c c c c c}
   \hline \hline
     & & FM  & & & &  AFM1 &  \\
   \cline{1-6}  \cline{7-8} 
   Config.  & Energy   & Cr$^{+3}$-Cr$^{+3}$ & Cr$^{+3}$-V$^{+5}$ &V$^{+5}$-V$^{+5}$ & & Energy  & \\
    &  (eV) & (\AA{}) & (\AA{}) & (\AA{}) & & (eV) & \\
   \hline
    1 & 0.0 & 4.450 & 3.433 & 4.450 & &+0.204  &\\
    2 & +0.155 & 5.902 & 2.962  3.461& 5.403 & & +0.400& \\
    3 & +0.019 & 3.486 & 2.962/3.461 & 5.316 &  &+0.224  &\\
    4 & +0.119  & 2.921 & 3.475 & 4.451 &  &+0.298 &\\
    5 & +0.064  & 5.341& 3.450& 4.452&  &+0.245  &\\
    6 & +0.017   &5.400 & 2.955/3.436 & 3.521 & &+0.302 &\\ 
    \hline \hline
  \end{tabular}
   \caption{Relative energies of different choice of FM and AFM1 configurations (columns 2 and 6) considered in a $\sqrt(2){\bf a} \times \sqrt(2) {\bf b} \times 4 {\bf c}$ supercell along with shortest distances  between different pairs of  Cr/V ions in columns 3, 4 and 5  are listed for the ferromagnetic configuration.}     
    \label{ch_CrVO2_tab.3}
\end{table}

\begin{figure}[ht]
\centering
\includegraphics[angle=90,height=10.0cm,width=16.0cm]{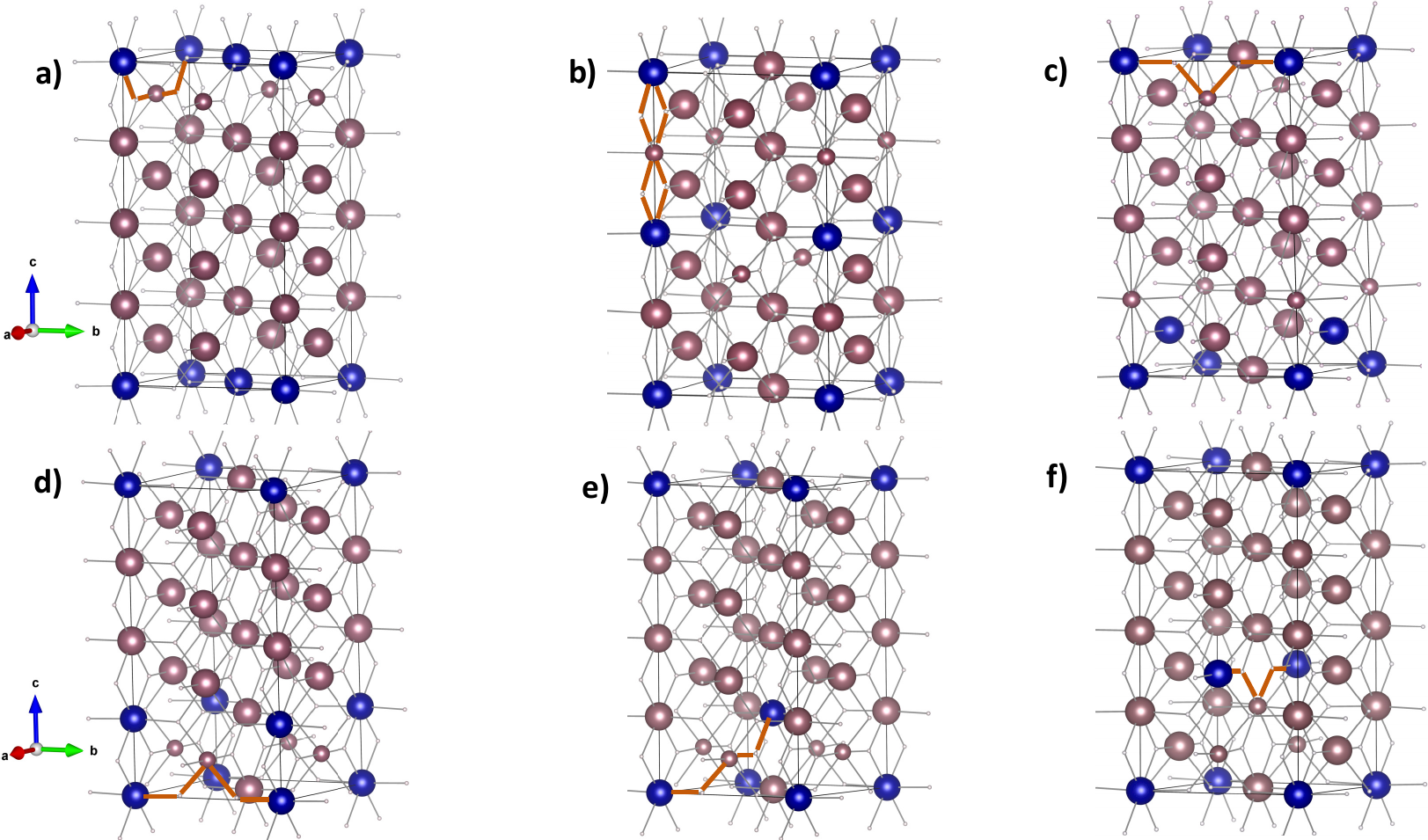}
\caption{Different choices considered for ferromagnetic arrangements of Cr$^{+3}$ (Blue balls), V$^{+4}$ (Big dark maroon balls) and V$^{+5}$ (Small dark maroon balls) in $\sqrt(2) {\bf a} \times \sqrt(2){\bf b} \times 4 {\bf c}$ supercell of 12.5\% Cr doped case. Small gray balls connecting these atoms are oxygen atoms.   One of the Cr- V$^{5+}$- Cr pathway is highlighted by thick saffron lines 
in each of these figures.}
  \label{fig.4}
\end{figure}

The choice of the supercell in the present instance has the V$^{+5}$ and Cr$^{+3}$ atoms arranged alternately in the $c$ direction along one 
of the chains. As this corresponds to a special configuration, we examined a larger unit cell of $\sqrt{2}{\bf a} \times \sqrt{2}{\bf b} \times 4{\bf c}$ which has 16 V atoms. We replaced two of the V atoms with Cr and explored all unique pairs possible within this unit cell. Relative energies with respect to the lowest ferromagnetic arrangement among various arrangement of doped Cr atoms in several ferromagnetic and possible AFM1 configurations are given in Table ~\ref{ch_CrVO2_tab.3}. We have also tabulated the shortest distance between the Cr$^{+3}$-Cr$^{+3}$, Cr$^{+3}$-V$^{+5}$ and V$^{+5}$-V$^{+5}$ pairs in the ferromagnetic case. In six out of the seven arrangement of doped Cr-atoms shown in Fig ~\ref{fig.4}, we have the ferromagnetic 
insulating state having lower energy. A continuous V$^{+5}$-Cr$^{+3}$ path is not necessary, but the presence of a pathway between two Cr atoms involving one V$^{+5}$ atom is found to stabilize the ferromagnetic insulating state. However, when the Cr atoms are far separated as in the seventh configuration, there is no such pathway possible. As a result, the ferromagnetic configuration is not favoured. Having established the nature of the ground state, we revert back to the 
smaller unit cell for further analysis to understand the microscopic considerations that determine this
unusual ground state.

\begin{figure}[ht]
\centering
\includegraphics[height=10.0cm,width=12.0cm]{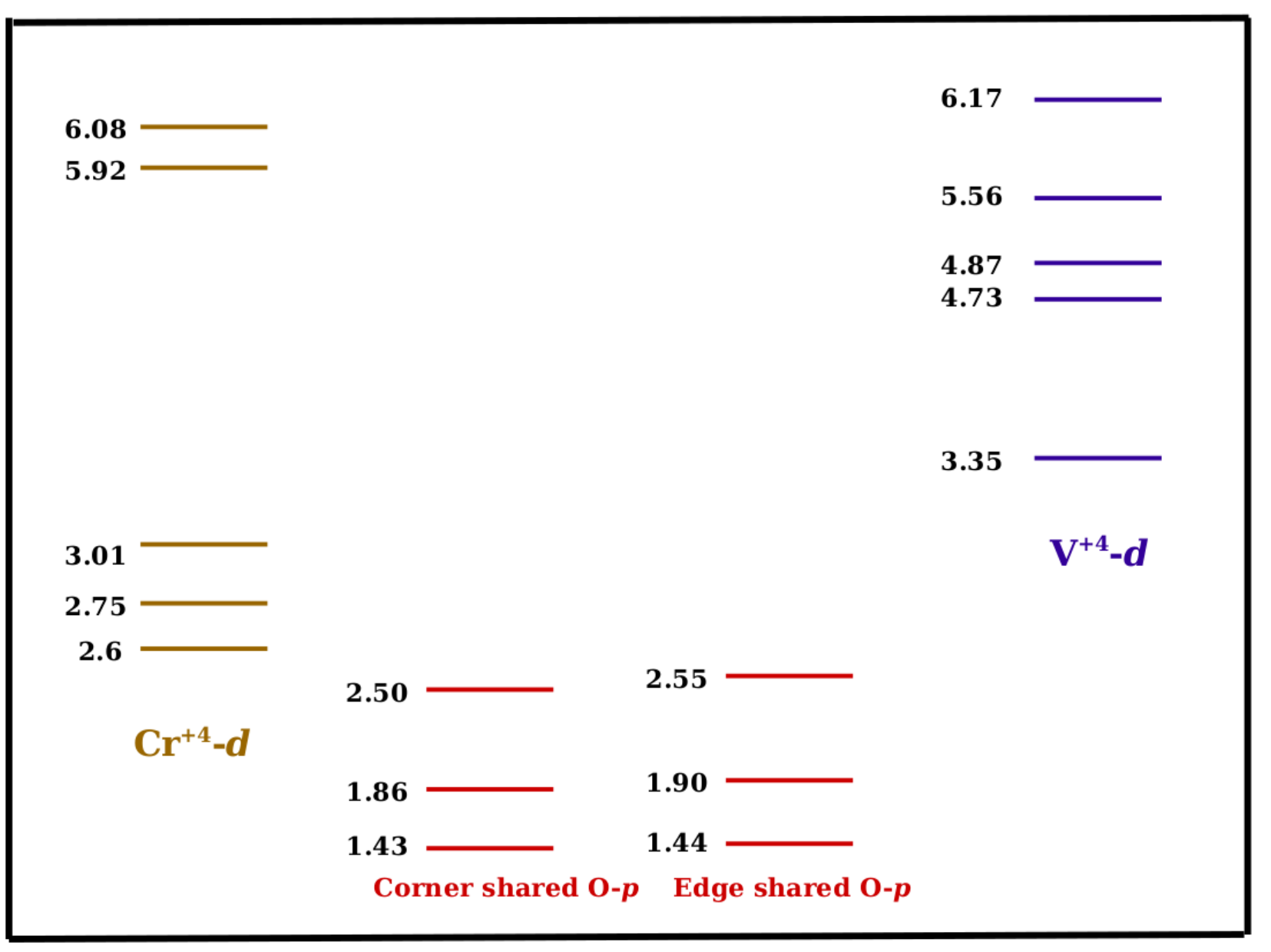}
\caption{Energy level diagram for Cr$^{+4}$ $d$ orbitals, corner and edge shared oxygen $p$ orbitals and V$^{+4}$ $d$ orbitals in the unrelaxed 
ferromagnetic structure for 12.5 \%  Cr doped VO$_2$. Energies (eV) for the majority spin channel have been given. 
This was obtained by fitting the $ab$ $initio$ band structure within a tight binding model considering Cr, V $d$ and O $p$ orbitals in the basis. The radial part of the tight binding basis functions are maximally localized Wannier functions.}
  \label{fig.5}
\end{figure}

We have calculated the {\it ab-initio} band structure along various 
symmetry directions for a 24  atom unit cell of  VO$_2$ in which one V atom is replaced by a Cr atom corresponding to a doping of 12.5 \%. No relaxation of the internal coordinates  was carried out.  A ferromagnetic configuration is assumed and a mapping onto a tight binding model was carried out for a model which had the transition metal $d$ and oxygen $p$ states in the basis. Here, the
radial parts of the wavefunctions are considered to be maximally localized Wannier functions. 
The onsite energies were extracted from the above mapping. Based upon this analysis, an energy 
level diagram showing the position of various orbitals in the majority spin channel is shown in Fig.~\ref{fig.5}.
Here we found that the lowest lying V-$d$ $t_{2g}$ orbital with the onsite energy 3.35 eV was $\sim$ 0.35 eV higher in
energy than the highest Cr-$d$ $t_{2g}$ level with the onsite energy $\sim$ 3.0 eV. This is the reason that the Cr $d$ levels have three electrons on them. 
This large stability for the Cr$^{+3}$ configuration emerges from the large Hund's stability associated with a half-filled $t_{2g}$ band. 
Allowing for an optimization of the structure, we find that the structural distortions aid 
this ordering at the Cr$^{+3}$ site. Additionally, one finds that the oxygens around one V have 
distorted resulting in a V$^{+5}$ configuration. The other V atoms are in 
the +4 configuration and the system becomes insulating after the relaxations. 

\begin{figure}[ht]
\centering
\includegraphics[height=6.0cm,width=13.0cm]{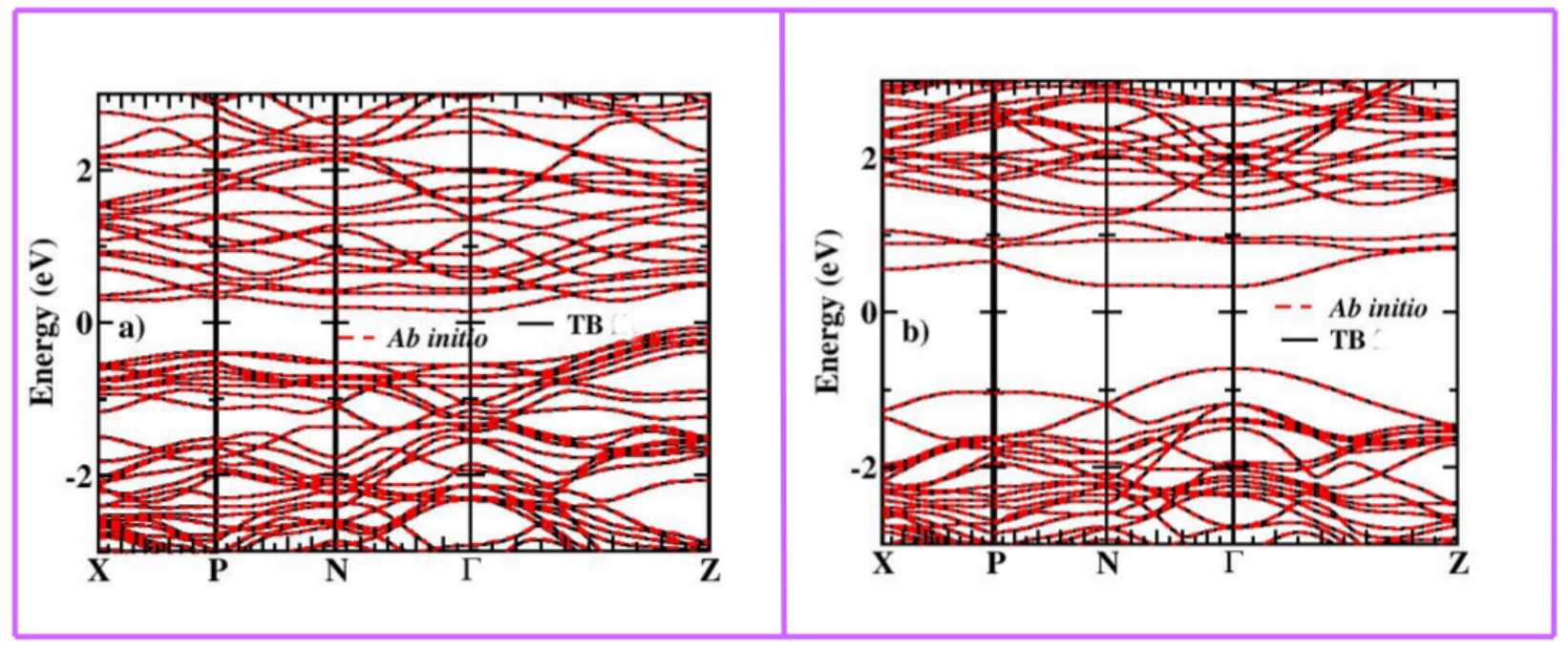}
\caption{A comparison of the $ab$ $initio$ spin-polarized band structure 
for ferromagnetic 12.5\% Cr doped VO$_2$ for (a) up and (b) down spin channel and that obtained within a  tight binding 
model considering Cr $d$, V $d$ and O $p$ states in the basis. 
Here, the radial part of the tight binding basis functions are maximally localized Wannier functions.}
  \label{fig.6}
\end{figure}

This analysis offers answers to few of the questions raised earlier, however few questions remain unanswered like why do we have a ferromagnetic ground state? 
Does the Cr$^{+3}$-V$^{+5}$ pair formation help in realizing a ferromagnetic state? 
In order to develop a clear understanding of these issues, we again carried out
a mapping of the {\it ab-initio} band structure for the ferromagnetic as well as the lowest
lying antiferromagnetic state onto a tight binding model at the same doping concentration. 
A comparison of the 
spin-polarized $ab$ $initio$ band structure with the tight binding one in both the spin channels is shown in Fig.~\ref{fig.6}. 
One finds that one has a reasonably good description of the band structure which gives us the confidence to use the Hamiltonian for further analysis. 
 
An estimate of the energy gain obtained from various hopping pathways available in this doped system would enable us to quantify their role  
in stabilization of such an unusual ground state. For this purpose, we artificially switch off the Cr$^{+3}$-V$^{+5}$ interactions 
via the intermediate oxygen atoms and calculated 
the band energy. A comparison of the evaluated band energy with the value obtained without switching off any interactions would provide us with the energy gain via the 
mentioned hopping pathways. This was done for both the ferromagnetic configuration as well as AFM1 which was the lowest lying antiferromagnetic configuration.
We found that the gain in energy via this hopping channel in the ferromagnetic configuration was $\sim$ 7 eV per hopping pathway.  
However, when a similar analysis was done after switching off direct interaction between Cr$^{+3}$-V$^{+5}$ pairs, we found a very small energy gain of 0.065 eV.
Based on these two results one can conclude that there is no direct pathway between the Cr$^{+3}$-V$^{+5}$ pair. The exchange pathway is through the oxygen atoms.  
We also estimated the energy gain from all Cr$^{+3}$-V$^{+4}$ hopping pathways via the oxygen atoms. 
We found the gain for the ferromagnetic configuration to be 3.11 eV per hopping pathway.  
We then did the same analysis for the next competing AFM1 configuration to present a comparative analysis in these two cases. The band energy gain 
in the AFM1 configuration from the most dominating channel of Cr$^{+3}$-V$^{+5}$ pairs via the oxygen atoms was found to be 6.293 eV/per hopping pathway. It can 
hence be clearly seen that this gain is $~$0.7 eV larger in the case of the ferromagnetic configuration when compared to 
AFM1 configuration. This helps in stablization of the 
ferromagnetic ground state. Contributions from direct interaction between Cr$^{+3}$-V$^{+5}$ pairs and from all Cr$^{+3}$-V$^{+4}$ hopping 
pathways via the oxygen atoms in the case of 
AFM1 are 0.217 eV and and 2.131 eV per hopping pathway respectively. Although the direct hopping
between Cr$^{+3}$-V$^{+4}$ pairs is larger in the AFM1 case, it is still 
substantially smaller than the channel via the oxygen atoms. The Cr$^{3+}$-V$^{+5}$ hopping interactions are again smaller in the 
AFM1 case when compared to the FM case.
The substantial gain from the exchange pathway between Cr-V sites suggests an important role played by the charge ordering in stabilizing a ferromagnetic 
insulating state. Similar ideas were discussed by some of us in the context of another ferromagnetic 
insulator K$_2$Cr$_8$O$_{16}$~\cite{kcro_th} and the present analysis for the 
first time is able to quantify the role of various exchange pathways. 
 
Previous $ab$ $initio$ study suggested a half metallic ferromagnetic character of Cr doped VO$_2$ in the rutile phase at 25\% doping~\cite{william6}, while higher doping concentrations closer to the CrO$_2$ end were examined in other theoretical papers \cite{william7,musto}. This study was done 
considering a 12 atom 
supercell where replacing one of the V with Cr atom corresponds to 25 \% doping. In order to investigate the system at this doping concentration, we replace two equivalent positions of the eight V atoms 
in our 24 atom supercell which corresponds to the same doping concentration. We allowed for volume optimization of this structure considering various magnetic configurations. 
The results comparing the total energy and and the shortest distance between the Cr$^{+3}$, Cr$^{+3}$-V$^{+5}$ and V$^{+5}$-V$^{+5}$  pairs different arrangement of Cr atoms in different ferromagnetic as well for various antiferromagnetic configurations at 25 \% doping are given in 
Table~S2 in the supplementary material~\cite{footnote-SI}, In Fig. S3, we have shown the arrangement of Cr$^{+3}$ and V$^{+4}$/V$^{+5}$ ions  in the lowest energy ferromagnetic FM3 structure along with all the other antiferromagnetic configurations considered. Similar to 12.5 \% case, we also find a ferromagnetic insulating ground state in this case with formation of Cr$^{+3}$-V$^{+5}$ pairs. 
Cr has a magnetic moment of 2.863 $\mu_B$ while the 
atoms identified as V$^{+4}$/V$^{+5}$ have magnetic moments 1.12/0.453 $\mu_B$ for FM3 configuration. We could not compare our results of 25 \% Cr doped case with experiments as no data is available at this doping concentration.

\section{Conclusion}
The ferromagnetic insulating state found in Cr doped VO$_2$ in the doping range from 10\% to 20\%, has been investigated at the concentration 
of 12.5\% within $ab$ $initio$ electronic structure calculations coupled with microscopic modeling within 
a tight binding model. We are able to reproduce the ferromagnetic 
insulating ground state within our calculations. Its stability is explained from the 
emergence of Cr$^{+3}$-V$^{+5}$ pairs which are found to be 
strongly stabilized by attractive Coulomb interactions. The presence of unoccupied $t_{2g}$ levels at the V sites
and occupied $t_{2g}$ levels at the Cr$^{+4}$ site leads 
to a superexchange pathway via the oxygens. This is found to strongly stabilize the ferromagnetic insulating state and its role is quantified for the first time.

%

\newpage

\bibliography{Vcro2}

\begin{thebibliography}{47}%
\makeatletter
\providecommand \@ifxundefined [1]{%
 \@ifx{#1\undefined}
}%
\providecommand \@ifnum [1]{%
 \ifnum #1\expandafter \@firstoftwo
 \else \expandafter \@secondoftwo
 \fi
}%
\providecommand \@ifx [1]{%
 \ifx #1\expandafter \@firstoftwo
 \else \expandafter \@secondoftwo
 \fi
}%
\providecommand \natexlab [1]{#1}%
\providecommand \enquote  [1]{``#1''}%
\providecommand \bibnamefont  [1]{#1}%
\providecommand \bibfnamefont [1]{#1}%
\providecommand \citenamefont [1]{#1}%
\providecommand \href@noop [0]{\@secondoftwo}%
\providecommand \href [0]{\begingroup \@sanitize@url \@href}%
\providecommand \@href[1]{\@@startlink{#1}\@@href}%
\providecommand \@@href[1]{\endgroup#1\@@endlink}%
\providecommand \@sanitize@url [0]{\catcode `\\12\catcode `\$12\catcode
  `\&12\catcode `\#12\catcode `\^12\catcode `\_12\catcode `\%12\relax}%
\providecommand \@@startlink[1]{}%
\providecommand \@@endlink[0]{}%
\providecommand \url  [0]{\begingroup\@sanitize@url \@url }%
\providecommand \@url [1]{\endgroup\@href {#1}{\urlprefix }}%
\providecommand \urlprefix  [0]{URL }%
\providecommand \Eprint [0]{\href }%
\providecommand \doibase [0]{http://dx.doi.org/}%
\providecommand \selectlanguage [0]{\@gobble}%
\providecommand \bibinfo  [0]{\@secondoftwo}%
\providecommand \bibfield  [0]{\@secondoftwo}%
\providecommand \translation [1]{[#1]}%
\providecommand \BibitemOpen [0]{}%
\providecommand \bibitemStop [0]{}%
\providecommand \bibitemNoStop [0]{.\EOS\space}%
\providecommand \EOS [0]{\spacefactor3000\relax}%
\providecommand \BibitemShut  [1]{\csname bibitem#1\endcsname}%
\let\auto@bib@innerbib\@empty
\bibitem [{\citenamefont {Chumak}\ \emph {et~al.}(2015)\citenamefont {Chumak},
  \citenamefont {Vasyuchka}, \citenamefont {Serga},\ and\ \citenamefont
  {Hillebrands}}]{fmi_appl1}%
  \BibitemOpen
  \bibfield  {author} {\bibinfo {author} {\bibfnamefont {A.~V.}\ \bibnamefont
  {Chumak}}, \bibinfo {author} {\bibfnamefont {V.~I.}\ \bibnamefont
  {Vasyuchka}}, \bibinfo {author} {\bibfnamefont {A.~A.}\ \bibnamefont
  {Serga}}, \ and\ \bibinfo {author} {\bibfnamefont {B.}~\bibnamefont
  {Hillebrands}},\ }\href {\doibase 10.1038/nphys3347} {\bibfield  {journal}
  {\bibinfo  {journal} {Nature Physics}\ }\textbf {\bibinfo {volume} {11}},\
  \bibinfo {pages} {453} (\bibinfo {year} {2015})}\BibitemShut {NoStop}%
\bibitem [{\citenamefont {Serga}\ \emph {et~al.}(2010)\citenamefont {Serga},
  \citenamefont {Chumak},\ and\ \citenamefont {Hillebrands}}]{fmi_appl2}%
  \BibitemOpen
  \bibfield  {author} {\bibinfo {author} {\bibfnamefont {A.~A.}\ \bibnamefont
  {Serga}}, \bibinfo {author} {\bibfnamefont {A.~V.}\ \bibnamefont {Chumak}}, \
  and\ \bibinfo {author} {\bibfnamefont {B.}~\bibnamefont {Hillebrands}},\
  }\href {\doibase 10.1088/0022-3727/43/26/264002} {\bibfield  {journal}
  {\bibinfo  {journal} {Journal of Physics D: Applied Physics}\ }\textbf
  {\bibinfo {volume} {43}},\ \bibinfo {pages} {264002} (\bibinfo {year}
  {2010})}\BibitemShut {NoStop}%
\bibitem [{\citenamefont {Huang}\ \emph {et~al.}(2017)\citenamefont {Huang},
  \citenamefont {Clark}, \citenamefont {Navarro-Moratalla}, \citenamefont
  {Klein}, \citenamefont {Cheng}, \citenamefont {Seyler}, \citenamefont
  {Zhong}, \citenamefont {Schmidgall}, \citenamefont {McGuire}, \citenamefont
  {Cobden}, \citenamefont {Yao}, \citenamefont {Xiao}, \citenamefont
  {Jarillo-Herrero},\ and\ \citenamefont {Xu}}]{cri3}%
  \BibitemOpen
  \bibfield  {author} {\bibinfo {author} {\bibfnamefont {B.}~\bibnamefont
  {Huang}}, \bibinfo {author} {\bibfnamefont {G.}~\bibnamefont {Clark}},
  \bibinfo {author} {\bibfnamefont {E.}~\bibnamefont {Navarro-Moratalla}},
  \bibinfo {author} {\bibfnamefont {D.~R.}\ \bibnamefont {Klein}}, \bibinfo
  {author} {\bibfnamefont {R.}~\bibnamefont {Cheng}}, \bibinfo {author}
  {\bibfnamefont {K.~L.}\ \bibnamefont {Seyler}}, \bibinfo {author}
  {\bibfnamefont {D.}~\bibnamefont {Zhong}}, \bibinfo {author} {\bibfnamefont
  {E.}~\bibnamefont {Schmidgall}}, \bibinfo {author} {\bibfnamefont {M.~A.}\
  \bibnamefont {McGuire}}, \bibinfo {author} {\bibfnamefont {D.~H.}\
  \bibnamefont {Cobden}}, \bibinfo {author} {\bibfnamefont {W.}~\bibnamefont
  {Yao}}, \bibinfo {author} {\bibfnamefont {D.}~\bibnamefont {Xiao}}, \bibinfo
  {author} {\bibfnamefont {P.}~\bibnamefont {Jarillo-Herrero}}, \ and\ \bibinfo
  {author} {\bibfnamefont {X.}~\bibnamefont {Xu}},\ }\href
  {https://doi.org/10.1038/nature22391} {\bibfield  {journal} {\bibinfo
  {journal} {Nature}\ }\textbf {\bibinfo {volume} {546}},\ \bibinfo {pages}
  {270} (\bibinfo {year} {2017})}\BibitemShut {NoStop}%
\bibitem [{\citenamefont {Tian}\ \emph {et~al.}(2019)\citenamefont {Tian},
  \citenamefont {Zhang}, \citenamefont {Li}, \citenamefont {Ying},
  \citenamefont {Li}, \citenamefont {Zhang}, \citenamefont {Liu},\ and\
  \citenamefont {Lei}}]{vi3}%
  \BibitemOpen
  \bibfield  {author} {\bibinfo {author} {\bibfnamefont {S.}~\bibnamefont
  {Tian}}, \bibinfo {author} {\bibfnamefont {J.-F.}\ \bibnamefont {Zhang}},
  \bibinfo {author} {\bibfnamefont {C.}~\bibnamefont {Li}}, \bibinfo {author}
  {\bibfnamefont {T.}~\bibnamefont {Ying}}, \bibinfo {author} {\bibfnamefont
  {S.}~\bibnamefont {Li}}, \bibinfo {author} {\bibfnamefont {X.}~\bibnamefont
  {Zhang}}, \bibinfo {author} {\bibfnamefont {K.}~\bibnamefont {Liu}}, \ and\
  \bibinfo {author} {\bibfnamefont {H.}~\bibnamefont {Lei}},\ }\href
  {https://doi.org/10.1021/jacs.8b13584} {\bibfield  {journal} {\bibinfo
  {journal} {J. Am. Chem. Soc.}\ }\textbf {\bibinfo {volume} {141}},\ \bibinfo
  {pages} {5326} (\bibinfo {year} {2019})}\BibitemShut {NoStop}%
\bibitem [{\citenamefont {Wakabayashi}\ \emph {et~al.}(2019)\citenamefont
  {Wakabayashi}, \citenamefont {Krockenberger}, \citenamefont {Tsujimoto},
  \citenamefont {Boykin}, \citenamefont {Tsuneyuki}, \citenamefont {Taniyasu},\
  and\ \citenamefont {Yamamoto}}]{sroso}%
  \BibitemOpen
  \bibfield  {author} {\bibinfo {author} {\bibfnamefont {Y.~K.}\ \bibnamefont
  {Wakabayashi}}, \bibinfo {author} {\bibfnamefont {Y.}~\bibnamefont
  {Krockenberger}}, \bibinfo {author} {\bibfnamefont {N.}~\bibnamefont
  {Tsujimoto}}, \bibinfo {author} {\bibfnamefont {T.}~\bibnamefont {Boykin}},
  \bibinfo {author} {\bibfnamefont {S.}~\bibnamefont {Tsuneyuki}}, \bibinfo
  {author} {\bibfnamefont {Y.}~\bibnamefont {Taniyasu}}, \ and\ \bibinfo
  {author} {\bibfnamefont {H.}~\bibnamefont {Yamamoto}},\ }\href
  {https://doi.org/10.1038/s41467-019-08440-6} {\bibfield  {journal} {\bibinfo
  {journal} {Nature Communications}\ }\textbf {\bibinfo {volume} {10}},\
  \bibinfo {pages} {535} (\bibinfo {year} {2019})}\BibitemShut {NoStop}%
\bibitem [{\citenamefont {Gong}\ \emph {et~al.}(2017)\citenamefont {Gong},
  \citenamefont {Li}, \citenamefont {Li}, \citenamefont {Ji}, \citenamefont
  {Stern}, \citenamefont {Xia}, \citenamefont {Cao}, \citenamefont {Bao},
  \citenamefont {Wang}, \citenamefont {Wang}, \citenamefont {Qiu},
  \citenamefont {Cava}, \citenamefont {Louie}, \citenamefont {Xia},\ and\
  \citenamefont {Zhang}}]{crgete}%
  \BibitemOpen
  \bibfield  {author} {\bibinfo {author} {\bibfnamefont {C.}~\bibnamefont
  {Gong}}, \bibinfo {author} {\bibfnamefont {L.}~\bibnamefont {Li}}, \bibinfo
  {author} {\bibfnamefont {Z.}~\bibnamefont {Li}}, \bibinfo {author}
  {\bibfnamefont {H.}~\bibnamefont {Ji}}, \bibinfo {author} {\bibfnamefont
  {A.}~\bibnamefont {Stern}}, \bibinfo {author} {\bibfnamefont
  {Y.}~\bibnamefont {Xia}}, \bibinfo {author} {\bibfnamefont {T.}~\bibnamefont
  {Cao}}, \bibinfo {author} {\bibfnamefont {W.}~\bibnamefont {Bao}}, \bibinfo
  {author} {\bibfnamefont {C.}~\bibnamefont {Wang}}, \bibinfo {author}
  {\bibfnamefont {Y.}~\bibnamefont {Wang}}, \bibinfo {author} {\bibfnamefont
  {Z.~Q.}\ \bibnamefont {Qiu}}, \bibinfo {author} {\bibfnamefont {R.~J.}\
  \bibnamefont {Cava}}, \bibinfo {author} {\bibfnamefont {S.~G.}\ \bibnamefont
  {Louie}}, \bibinfo {author} {\bibfnamefont {J.}~\bibnamefont {Xia}}, \ and\
  \bibinfo {author} {\bibfnamefont {X.}~\bibnamefont {Zhang}},\ }\href
  {https://doi.org/10.1038/nature22060} {\bibfield  {journal} {\bibinfo
  {journal} {Nature}\ }\textbf {\bibinfo {volume} {546}},\ \bibinfo {pages}
  {265} (\bibinfo {year} {2017})}\BibitemShut {NoStop}%
\bibitem [{\citenamefont {Hemberger}\ \emph {et~al.}(2005)\citenamefont
  {Hemberger}, \citenamefont {Lunkenheimer}, \citenamefont {R}, \citenamefont
  {von Nidda~HA}, \citenamefont {V},\ and\ \citenamefont {A}}]{cdcrs}%
  \BibitemOpen
  \bibfield  {author} {\bibinfo {author} {\bibfnamefont {J.}~\bibnamefont
  {Hemberger}}, \bibinfo {author} {\bibfnamefont {P.}~\bibnamefont
  {Lunkenheimer}}, \bibinfo {author} {\bibfnamefont {F.}~\bibnamefont {R}},
  \bibinfo {author} {\bibfnamefont {K.}~\bibnamefont {von Nidda~HA}}, \bibinfo
  {author} {\bibfnamefont {T.}~\bibnamefont {V}}, \ and\ \bibinfo {author}
  {\bibfnamefont {L.}~\bibnamefont {A}},\ }\href
  {https://doi.org/10.1038/nature22060} {\bibfield  {journal} {\bibinfo
  {journal} {Nature}\ }\textbf {\bibinfo {volume} {434}},\ \bibinfo {pages}
  {364} (\bibinfo {year} {2005})}\BibitemShut {NoStop}%
\bibitem [{\citenamefont {Azuma}\ \emph {et~al.}(2005)\citenamefont {Azuma},
  \citenamefont {Takata}, \citenamefont {Saito}, \citenamefont {Ishiwata},
  \citenamefont {Shimakawa},\ and\ \citenamefont {Takano}}]{bnmo}%
  \BibitemOpen
  \bibfield  {author} {\bibinfo {author} {\bibfnamefont {M.}~\bibnamefont
  {Azuma}}, \bibinfo {author} {\bibfnamefont {K.}~\bibnamefont {Takata}},
  \bibinfo {author} {\bibfnamefont {T.}~\bibnamefont {Saito}}, \bibinfo
  {author} {\bibfnamefont {S.}~\bibnamefont {Ishiwata}}, \bibinfo {author}
  {\bibfnamefont {Y.}~\bibnamefont {Shimakawa}}, \ and\ \bibinfo {author}
  {\bibfnamefont {M.}~\bibnamefont {Takano}},\ }\href
  {https://doi.org/10.1021/ja0512576} {\bibfield  {journal} {\bibinfo
  {journal} {J. Am. Chem. Soc.}\ }\textbf {\bibinfo {volume} {127}},\ \bibinfo
  {pages} {8889} (\bibinfo {year} {2005})}\BibitemShut {NoStop}%
\bibitem [{\citenamefont {Das}\ \emph {et~al.}(2008)\citenamefont {Das},
  \citenamefont {Waghmare}, \citenamefont {Saha-Dasgupta},\ and\ \citenamefont
  {Sarma}}]{lnmo}%
  \BibitemOpen
  \bibfield  {author} {\bibinfo {author} {\bibfnamefont {H.}~\bibnamefont
  {Das}}, \bibinfo {author} {\bibfnamefont {U.~V.}\ \bibnamefont {Waghmare}},
  \bibinfo {author} {\bibfnamefont {T.}~\bibnamefont {Saha-Dasgupta}}, \ and\
  \bibinfo {author} {\bibfnamefont {D.~D.}\ \bibnamefont {Sarma}},\ }\href
  {\doibase 10.1103/PhysRevLett.100.186402} {\bibfield  {journal} {\bibinfo
  {journal} {Phys. Rev. Lett.}\ }\textbf {\bibinfo {volume} {100}},\ \bibinfo
  {pages} {186402} (\bibinfo {year} {2008})}\BibitemShut {NoStop}%
\bibitem [{\citenamefont {Pavarini}\ \emph {et~al.}(2005)\citenamefont
  {Pavarini}, \citenamefont {Yamasaki}, \citenamefont {Nuss},\ and\
  \citenamefont {Andersen}}]{ytio}%
  \BibitemOpen
  \bibfield  {author} {\bibinfo {author} {\bibfnamefont {E.}~\bibnamefont
  {Pavarini}}, \bibinfo {author} {\bibfnamefont {A.}~\bibnamefont {Yamasaki}},
  \bibinfo {author} {\bibfnamefont {J.}~\bibnamefont {Nuss}}, \ and\ \bibinfo
  {author} {\bibfnamefont {O.~K.}\ \bibnamefont {Andersen}},\ }\href {\doibase
  10.1088/1367-2630/7/1/188} {\bibfield  {journal} {\bibinfo  {journal} {New
  Journal of Physics}\ }\textbf {\bibinfo {volume} {7}},\ \bibinfo {pages}
  {188} (\bibinfo {year} {2005})}\BibitemShut {NoStop}%
\bibitem [{\citenamefont {Meng}\ \emph {et~al.}(2018)\citenamefont {Meng},
  \citenamefont {Guo}, \citenamefont {Cui}, \citenamefont {Ma}, \citenamefont
  {Zhao}, \citenamefont {Lu}, \citenamefont {Xu}, \citenamefont {Wang},
  \citenamefont {Hu}, \citenamefont {Fu}, \citenamefont {Peng}, \citenamefont
  {Guo}, \citenamefont {Zhai}, \citenamefont {Brown}, \citenamefont {Knize},\
  and\ \citenamefont {Lu}}]{lcoo}%
  \BibitemOpen
  \bibfield  {author} {\bibinfo {author} {\bibfnamefont {D.}~\bibnamefont
  {Meng}}, \bibinfo {author} {\bibfnamefont {H.}~\bibnamefont {Guo}}, \bibinfo
  {author} {\bibfnamefont {Z.}~\bibnamefont {Cui}}, \bibinfo {author}
  {\bibfnamefont {C.}~\bibnamefont {Ma}}, \bibinfo {author} {\bibfnamefont
  {J.}~\bibnamefont {Zhao}}, \bibinfo {author} {\bibfnamefont {J.}~\bibnamefont
  {Lu}}, \bibinfo {author} {\bibfnamefont {H.}~\bibnamefont {Xu}}, \bibinfo
  {author} {\bibfnamefont {Z.}~\bibnamefont {Wang}}, \bibinfo {author}
  {\bibfnamefont {X.}~\bibnamefont {Hu}}, \bibinfo {author} {\bibfnamefont
  {Z.}~\bibnamefont {Fu}}, \bibinfo {author} {\bibfnamefont {R.}~\bibnamefont
  {Peng}}, \bibinfo {author} {\bibfnamefont {J.}~\bibnamefont {Guo}}, \bibinfo
  {author} {\bibfnamefont {X.}~\bibnamefont {Zhai}}, \bibinfo {author}
  {\bibfnamefont {G.~J.}\ \bibnamefont {Brown}}, \bibinfo {author}
  {\bibfnamefont {R.}~\bibnamefont {Knize}}, \ and\ \bibinfo {author}
  {\bibfnamefont {Y.}~\bibnamefont {Lu}},\ }\href {\doibase
  10.1073/pnas.1707817115} {\bibfield  {journal} {\bibinfo  {journal}
  {Proceedings of the National Academy of Sciences}\ }\textbf {\bibinfo
  {volume} {115}},\ \bibinfo {pages} {2873} (\bibinfo {year} {2018})},\ \Eprint
  {http://arxiv.org/abs/https://www.pnas.org/content/115/12/2873.full.pdf}
  {https://www.pnas.org/content/115/12/2873.full.pdf} \BibitemShut {NoStop}%
\bibitem [{\citenamefont {Algarabel}\ \emph {et~al.}(2003)\citenamefont
  {Algarabel}, \citenamefont {De~Teresa}, \citenamefont {Blasco}, \citenamefont
  {Ibarra}, \citenamefont {Kapusta}, \citenamefont {Sikora}, \citenamefont
  {Zajac}, \citenamefont {Riedi},\ and\ \citenamefont {Ritter}}]{manganites}%
  \BibitemOpen
  \bibfield  {author} {\bibinfo {author} {\bibfnamefont {P.~A.}\ \bibnamefont
  {Algarabel}}, \bibinfo {author} {\bibfnamefont {J.~M.}\ \bibnamefont
  {De~Teresa}}, \bibinfo {author} {\bibfnamefont {J.}~\bibnamefont {Blasco}},
  \bibinfo {author} {\bibfnamefont {M.~R.}\ \bibnamefont {Ibarra}}, \bibinfo
  {author} {\bibfnamefont {C.}~\bibnamefont {Kapusta}}, \bibinfo {author}
  {\bibfnamefont {M.}~\bibnamefont {Sikora}}, \bibinfo {author} {\bibfnamefont
  {D.}~\bibnamefont {Zajac}}, \bibinfo {author} {\bibfnamefont {P.~C.}\
  \bibnamefont {Riedi}}, \ and\ \bibinfo {author} {\bibfnamefont
  {C.}~\bibnamefont {Ritter}},\ }\href {\doibase 10.1103/PhysRevB.67.134402}
  {\bibfield  {journal} {\bibinfo  {journal} {Phys. Rev. B}\ }\textbf {\bibinfo
  {volume} {67}},\ \bibinfo {pages} {134402} (\bibinfo {year}
  {2003})}\BibitemShut {NoStop}%
\bibitem [{\citenamefont {Hasegawa}\ \emph {et~al.}(2009)\citenamefont
  {Hasegawa}, \citenamefont {Isobe}, \citenamefont {Yamauchi}, \citenamefont
  {Ueda}, \citenamefont {Yamaura}, \citenamefont {Gotou}, \citenamefont {Yagi},
  \citenamefont {Sato},\ and\ \citenamefont {Ueda}}]{kcro_expt}%
  \BibitemOpen
  \bibfield  {author} {\bibinfo {author} {\bibfnamefont {K.}~\bibnamefont
  {Hasegawa}}, \bibinfo {author} {\bibfnamefont {M.}~\bibnamefont {Isobe}},
  \bibinfo {author} {\bibfnamefont {T.}~\bibnamefont {Yamauchi}}, \bibinfo
  {author} {\bibfnamefont {H.}~\bibnamefont {Ueda}}, \bibinfo {author}
  {\bibfnamefont {J.-I.}\ \bibnamefont {Yamaura}}, \bibinfo {author}
  {\bibfnamefont {H.}~\bibnamefont {Gotou}}, \bibinfo {author} {\bibfnamefont
  {T.}~\bibnamefont {Yagi}}, \bibinfo {author} {\bibfnamefont {H.}~\bibnamefont
  {Sato}}, \ and\ \bibinfo {author} {\bibfnamefont {Y.}~\bibnamefont {Ueda}},\
  }\href {\doibase 10.1103/PhysRevLett.103.146403} {\bibfield  {journal}
  {\bibinfo  {journal} {Phys. Rev. Lett.}\ }\textbf {\bibinfo {volume} {103}},\
  \bibinfo {pages} {146403} (\bibinfo {year} {2009})}\BibitemShut {NoStop}%
\bibitem [{\citenamefont {Mahadevan}\ \emph {et~al.}(2010)\citenamefont
  {Mahadevan}, \citenamefont {Kumar}, \citenamefont {Choudhury},\ and\
  \citenamefont {Sarma}}]{kcro_th}%
  \BibitemOpen
  \bibfield  {author} {\bibinfo {author} {\bibfnamefont {P.}~\bibnamefont
  {Mahadevan}}, \bibinfo {author} {\bibfnamefont {A.}~\bibnamefont {Kumar}},
  \bibinfo {author} {\bibfnamefont {D.}~\bibnamefont {Choudhury}}, \ and\
  \bibinfo {author} {\bibfnamefont {D.~D.}\ \bibnamefont {Sarma}},\ }\href
  {\doibase 10.1103/PhysRevLett.104.256401} {\bibfield  {journal} {\bibinfo
  {journal} {Phys. Rev. Lett.}\ }\textbf {\bibinfo {volume} {104}},\ \bibinfo
  {pages} {256401} (\bibinfo {year} {2010})}\BibitemShut {NoStop}%
\bibitem [{\citenamefont {Marezio}\ \emph {et~al.}(1972)\citenamefont
  {Marezio}, \citenamefont {McWhan}, \citenamefont {Remeika},\ and\
  \citenamefont {Dernier}}]{mono-26}%
  \BibitemOpen
  \bibfield  {author} {\bibinfo {author} {\bibfnamefont {M.}~\bibnamefont
  {Marezio}}, \bibinfo {author} {\bibfnamefont {D.~B.}\ \bibnamefont {McWhan}},
  \bibinfo {author} {\bibfnamefont {J.~P.}\ \bibnamefont {Remeika}}, \ and\
  \bibinfo {author} {\bibfnamefont {P.~D.}\ \bibnamefont {Dernier}},\ }\href
  {\doibase 10.1103/PhysRevB.5.2541} {\bibfield  {journal} {\bibinfo  {journal}
  {Phys. Rev. B}\ }\textbf {\bibinfo {volume} {5}},\ \bibinfo {pages} {2541}
  (\bibinfo {year} {1972})}\BibitemShut {NoStop}%
\bibitem [{\citenamefont {Goodenough}\ and\ \citenamefont
  {Hong}(1973)}]{mono-16}%
  \BibitemOpen
  \bibfield  {author} {\bibinfo {author} {\bibfnamefont {J.~B.}\ \bibnamefont
  {Goodenough}}\ and\ \bibinfo {author} {\bibfnamefont {H.~Y.-P.}\ \bibnamefont
  {Hong}},\ }\href {\doibase 10.1103/PhysRevB.8.1323} {\bibfield  {journal}
  {\bibinfo  {journal} {Phys. Rev. B}\ }\textbf {\bibinfo {volume} {8}},\
  \bibinfo {pages} {1323} (\bibinfo {year} {1973})}\BibitemShut {NoStop}%
\bibitem [{\citenamefont {Pouget}\ \emph {et~al.}(1975)\citenamefont {Pouget},
  \citenamefont {Launois}, \citenamefont {D'Haenens}, \citenamefont {Merenda},\
  and\ \citenamefont {Rice}}]{poupet6}%
  \BibitemOpen
  \bibfield  {author} {\bibinfo {author} {\bibfnamefont {J.~P.}\ \bibnamefont
  {Pouget}}, \bibinfo {author} {\bibfnamefont {H.}~\bibnamefont {Launois}},
  \bibinfo {author} {\bibfnamefont {J.~P.}\ \bibnamefont {D'Haenens}}, \bibinfo
  {author} {\bibfnamefont {P.}~\bibnamefont {Merenda}}, \ and\ \bibinfo
  {author} {\bibfnamefont {T.~M.}\ \bibnamefont {Rice}},\ }\href {\doibase
  10.1103/PhysRevLett.35.873} {\bibfield  {journal} {\bibinfo  {journal} {Phys.
  Rev. Lett.}\ }\textbf {\bibinfo {volume} {35}},\ \bibinfo {pages} {873}
  (\bibinfo {year} {1975})}\BibitemShut {NoStop}%
\bibitem [{\citenamefont {West}\ \emph {et~al.}(2008)\citenamefont {West},
  \citenamefont {Lu}, \citenamefont {He}, \citenamefont {Kirkwood},
  \citenamefont {Chen}, \citenamefont {Adl}, \citenamefont {Osofsky},
  \citenamefont {Qadri}, \citenamefont {Hull},\ and\ \citenamefont
  {Wolf}}]{exp}%
  \BibitemOpen
  \bibfield  {author} {\bibinfo {author} {\bibfnamefont {K.~G.}\ \bibnamefont
  {West}}, \bibinfo {author} {\bibfnamefont {J.}~\bibnamefont {Lu}}, \bibinfo
  {author} {\bibfnamefont {L.}~\bibnamefont {He}}, \bibinfo {author}
  {\bibfnamefont {D.}~\bibnamefont {Kirkwood}}, \bibinfo {author}
  {\bibfnamefont {W.}~\bibnamefont {Chen}}, \bibinfo {author} {\bibfnamefont
  {T.~P.}\ \bibnamefont {Adl}}, \bibinfo {author} {\bibfnamefont {M.~S.}\
  \bibnamefont {Osofsky}}, \bibinfo {author} {\bibfnamefont {S.~B.}\
  \bibnamefont {Qadri}}, \bibinfo {author} {\bibfnamefont {R.}~\bibnamefont
  {Hull}}, \ and\ \bibinfo {author} {\bibfnamefont {S.~A.}\ \bibnamefont
  {Wolf}},\ }\href {\doibase 10.1007/s10948-007-0303-y} {\bibfield  {journal}
  {\bibinfo  {journal} {Journal of Superconductivity and Novel Magnetism}\
  }\textbf {\bibinfo {volume} {21}},\ \bibinfo {pages} {87} (\bibinfo {year}
  {2008})}\BibitemShut {NoStop}%
\bibitem [{\citenamefont {Piper}\ \emph {et~al.}(2010)\citenamefont {Piper},
  \citenamefont {DeMasi}, \citenamefont {Cho}, \citenamefont {Preston},
  \citenamefont {Laverock}, \citenamefont {Smith}, \citenamefont {West},
  \citenamefont {Lu},\ and\ \citenamefont {Wolf}}]{piper}%
  \BibitemOpen
  \bibfield  {author} {\bibinfo {author} {\bibfnamefont {L.~F.~J.}\
  \bibnamefont {Piper}}, \bibinfo {author} {\bibfnamefont {A.}~\bibnamefont
  {DeMasi}}, \bibinfo {author} {\bibfnamefont {S.~W.}\ \bibnamefont {Cho}},
  \bibinfo {author} {\bibfnamefont {A.~R.~H.}\ \bibnamefont {Preston}},
  \bibinfo {author} {\bibfnamefont {J.}~\bibnamefont {Laverock}}, \bibinfo
  {author} {\bibfnamefont {K.~E.}\ \bibnamefont {Smith}}, \bibinfo {author}
  {\bibfnamefont {K.~G.}\ \bibnamefont {West}}, \bibinfo {author}
  {\bibfnamefont {J.~W.}\ \bibnamefont {Lu}}, \ and\ \bibinfo {author}
  {\bibfnamefont {S.~A.}\ \bibnamefont {Wolf}},\ }\href {\doibase
  10.1103/PhysRevB.82.235103} {\bibfield  {journal} {\bibinfo  {journal} {Phys.
  Rev. B}\ }\textbf {\bibinfo {volume} {82}},\ \bibinfo {pages} {235103}
  (\bibinfo {year} {2010})}\BibitemShut {NoStop}%
\bibitem [{\citenamefont {Yang}\ \emph {et~al.}(2011)\citenamefont {Yang},
  \citenamefont {Ko},\ and\ \citenamefont {Ramanathan}}]{vo2_exp1}%
  \BibitemOpen
  \bibfield  {author} {\bibinfo {author} {\bibfnamefont {Z.}~\bibnamefont
  {Yang}}, \bibinfo {author} {\bibfnamefont {C.}~\bibnamefont {Ko}}, \ and\
  \bibinfo {author} {\bibfnamefont {S.}~\bibnamefont {Ramanathan}},\ }\href
  {\doibase 10.1146/annurev-matsci-062910-100347} {\bibfield  {journal}
  {\bibinfo  {journal} {Annual Review of Materials Research}\ }\textbf
  {\bibinfo {volume} {41}},\ \bibinfo {pages} {337} (\bibinfo {year}
  {2011})}\BibitemShut {NoStop}%
\bibitem [{\citenamefont {Morin}(1959)}]{tc6}%
  \BibitemOpen
  \bibfield  {author} {\bibinfo {author} {\bibfnamefont {F.~J.}\ \bibnamefont
  {Morin}},\ }\href {\doibase 10.1103/PhysRevLett.3.34} {\bibfield  {journal}
  {\bibinfo  {journal} {Phys. Rev. Lett.}\ }\textbf {\bibinfo {volume} {3}},\
  \bibinfo {pages} {34} (\bibinfo {year} {1959})}\BibitemShut {NoStop}%
\bibitem [{\citenamefont {Haverkort}\ \emph {et~al.}(2005)\citenamefont
  {Haverkort}, \citenamefont {Hu}, \citenamefont {Tanaka}, \citenamefont
  {Reichelt}, \citenamefont {Streltsov}, \citenamefont {Korotin}, \citenamefont
  {Anisimov}, \citenamefont {Hsieh}, \citenamefont {Lin}, \citenamefont {Chen},
  \citenamefont {Khomskii},\ and\ \citenamefont {Tjeng}}]{mot6}%
  \BibitemOpen
  \bibfield  {author} {\bibinfo {author} {\bibfnamefont {M.~W.}\ \bibnamefont
  {Haverkort}}, \bibinfo {author} {\bibfnamefont {Z.}~\bibnamefont {Hu}},
  \bibinfo {author} {\bibfnamefont {A.}~\bibnamefont {Tanaka}}, \bibinfo
  {author} {\bibfnamefont {W.}~\bibnamefont {Reichelt}}, \bibinfo {author}
  {\bibfnamefont {S.~V.}\ \bibnamefont {Streltsov}}, \bibinfo {author}
  {\bibfnamefont {M.~A.}\ \bibnamefont {Korotin}}, \bibinfo {author}
  {\bibfnamefont {V.~I.}\ \bibnamefont {Anisimov}}, \bibinfo {author}
  {\bibfnamefont {H.~H.}\ \bibnamefont {Hsieh}}, \bibinfo {author}
  {\bibfnamefont {H.-J.}\ \bibnamefont {Lin}}, \bibinfo {author} {\bibfnamefont
  {C.~T.}\ \bibnamefont {Chen}}, \bibinfo {author} {\bibfnamefont {D.~I.}\
  \bibnamefont {Khomskii}}, \ and\ \bibinfo {author} {\bibfnamefont {L.~H.}\
  \bibnamefont {Tjeng}},\ }\href {\doibase 10.1103/PhysRevLett.95.196404}
  {\bibfield  {journal} {\bibinfo  {journal} {Phys. Rev. Lett.}\ }\textbf
  {\bibinfo {volume} {95}},\ \bibinfo {pages} {196404} (\bibinfo {year}
  {2005})}\BibitemShut {NoStop}%
\bibitem [{\citenamefont {Koethe}\ \emph {et~al.}(2006)\citenamefont {Koethe},
  \citenamefont {Hu}, \citenamefont {Haverkort}, \citenamefont
  {Sch\"u\ss{}ler-Langeheine}, \citenamefont {Venturini}, \citenamefont
  {Brookes}, \citenamefont {Tjernberg}, \citenamefont {Reichelt}, \citenamefont
  {Hsieh}, \citenamefont {Lin}, \citenamefont {Chen},\ and\ \citenamefont
  {Tjeng}}]{peierl6}%
  \BibitemOpen
  \bibfield  {author} {\bibinfo {author} {\bibfnamefont {T.~C.}\ \bibnamefont
  {Koethe}}, \bibinfo {author} {\bibfnamefont {Z.}~\bibnamefont {Hu}}, \bibinfo
  {author} {\bibfnamefont {M.~W.}\ \bibnamefont {Haverkort}}, \bibinfo {author}
  {\bibfnamefont {C.}~\bibnamefont {Sch\"u\ss{}ler-Langeheine}}, \bibinfo
  {author} {\bibfnamefont {F.}~\bibnamefont {Venturini}}, \bibinfo {author}
  {\bibfnamefont {N.~B.}\ \bibnamefont {Brookes}}, \bibinfo {author}
  {\bibfnamefont {O.}~\bibnamefont {Tjernberg}}, \bibinfo {author}
  {\bibfnamefont {W.}~\bibnamefont {Reichelt}}, \bibinfo {author}
  {\bibfnamefont {H.~H.}\ \bibnamefont {Hsieh}}, \bibinfo {author}
  {\bibfnamefont {H.-J.}\ \bibnamefont {Lin}}, \bibinfo {author} {\bibfnamefont
  {C.~T.}\ \bibnamefont {Chen}}, \ and\ \bibinfo {author} {\bibfnamefont
  {L.~H.}\ \bibnamefont {Tjeng}},\ }\href {\doibase
  10.1103/PhysRevLett.97.116402} {\bibfield  {journal} {\bibinfo  {journal}
  {Phys. Rev. Lett.}\ }\textbf {\bibinfo {volume} {97}},\ \bibinfo {pages}
  {116402} (\bibinfo {year} {2006})}\BibitemShut {NoStop}%
\bibitem [{\citenamefont {Soulen}\ \emph {et~al.}(1998)\citenamefont {Soulen},
  \citenamefont {Byers}, \citenamefont {Osofsky}, \citenamefont {Nadgorny},
  \citenamefont {Ambrose}, \citenamefont {Cheng}, \citenamefont {Broussard},
  \citenamefont {Tanaka}, \citenamefont {Nowak}, \citenamefont {Moodera},
  \citenamefont {Barry},\ and\ \citenamefont {Coey}}]{cro2_exp1}%
  \BibitemOpen
  \bibfield  {author} {\bibinfo {author} {\bibfnamefont {R.~J.}\ \bibnamefont
  {Soulen}}, \bibinfo {author} {\bibfnamefont {J.~M.}\ \bibnamefont {Byers}},
  \bibinfo {author} {\bibfnamefont {M.~S.}\ \bibnamefont {Osofsky}}, \bibinfo
  {author} {\bibfnamefont {B.}~\bibnamefont {Nadgorny}}, \bibinfo {author}
  {\bibfnamefont {T.}~\bibnamefont {Ambrose}}, \bibinfo {author} {\bibfnamefont
  {S.~F.}\ \bibnamefont {Cheng}}, \bibinfo {author} {\bibfnamefont {P.~R.}\
  \bibnamefont {Broussard}}, \bibinfo {author} {\bibfnamefont {C.~T.}\
  \bibnamefont {Tanaka}}, \bibinfo {author} {\bibfnamefont {J.}~\bibnamefont
  {Nowak}}, \bibinfo {author} {\bibfnamefont {J.~S.}\ \bibnamefont {Moodera}},
  \bibinfo {author} {\bibfnamefont {A.}~\bibnamefont {Barry}}, \ and\ \bibinfo
  {author} {\bibfnamefont {J.~M.~D.}\ \bibnamefont {Coey}},\ }\href {\doibase
  10.1126/science.282.5386.85} {\bibfield  {journal} {\bibinfo  {journal}
  {Science}\ }\textbf {\bibinfo {volume} {282}},\ \bibinfo {pages} {85}
  (\bibinfo {year} {1998})}\BibitemShut {NoStop}%
\bibitem [{\citenamefont {Chamberland}(1977)}]{cro2_exp2}%
  \BibitemOpen
  \bibfield  {author} {\bibinfo {author} {\bibfnamefont {B.~L.}\ \bibnamefont
  {Chamberland}},\ }\href {\doibase 10.1080/10408437708243431} {\bibfield
  {journal} {\bibinfo  {journal} {Critical Reviews in Solid State and Materials
  Sciences}\ }\textbf {\bibinfo {volume} {7}},\ \bibinfo {pages} {1} (\bibinfo
  {year} {1977})}\BibitemShut {NoStop}%
\bibitem [{\citenamefont {Dedkov}\ \emph {et~al.}(2005)\citenamefont {Dedkov},
  \citenamefont {Vinogradov}, \citenamefont {Fonin}, \citenamefont {K\"onig},
  \citenamefont {Vyalikh}, \citenamefont {Preobrajenski}, \citenamefont
  {Krasnikov}, \citenamefont {Kleimenov}, \citenamefont {Nesterov},
  \citenamefont {R\"udiger}, \citenamefont {Molodtsov},\ and\ \citenamefont
  {G\"untherodt}}]{cro2_exp3}%
  \BibitemOpen
  \bibfield  {author} {\bibinfo {author} {\bibfnamefont {Y.~S.}\ \bibnamefont
  {Dedkov}}, \bibinfo {author} {\bibfnamefont {A.~S.}\ \bibnamefont
  {Vinogradov}}, \bibinfo {author} {\bibfnamefont {M.}~\bibnamefont {Fonin}},
  \bibinfo {author} {\bibfnamefont {C.}~\bibnamefont {K\"onig}}, \bibinfo
  {author} {\bibfnamefont {D.~V.}\ \bibnamefont {Vyalikh}}, \bibinfo {author}
  {\bibfnamefont {A.~B.}\ \bibnamefont {Preobrajenski}}, \bibinfo {author}
  {\bibfnamefont {S.~A.}\ \bibnamefont {Krasnikov}}, \bibinfo {author}
  {\bibfnamefont {E.~Y.}\ \bibnamefont {Kleimenov}}, \bibinfo {author}
  {\bibfnamefont {M.~A.}\ \bibnamefont {Nesterov}}, \bibinfo {author}
  {\bibfnamefont {U.}~\bibnamefont {R\"udiger}}, \bibinfo {author}
  {\bibfnamefont {S.~L.}\ \bibnamefont {Molodtsov}}, \ and\ \bibinfo {author}
  {\bibfnamefont {G.}~\bibnamefont {G\"untherodt}},\ }\href {\doibase
  10.1103/PhysRevB.72.060401} {\bibfield  {journal} {\bibinfo  {journal} {Phys.
  Rev. B}\ }\textbf {\bibinfo {volume} {72}},\ \bibinfo {pages} {060401}
  (\bibinfo {year} {2005})}\BibitemShut {NoStop}%
\bibitem [{\citenamefont {Coey}\ \emph {et~al.}(1998)\citenamefont {Coey},
  \citenamefont {Berkowitz}, \citenamefont {Balcells}, \citenamefont {Putris},\
  and\ \citenamefont {Barry}}]{cro2_exp4}%
  \BibitemOpen
  \bibfield  {author} {\bibinfo {author} {\bibfnamefont {J.~M.~D.}\
  \bibnamefont {Coey}}, \bibinfo {author} {\bibfnamefont {A.~E.}\ \bibnamefont
  {Berkowitz}}, \bibinfo {author} {\bibfnamefont {L.}~\bibnamefont {Balcells}},
  \bibinfo {author} {\bibfnamefont {F.~F.}\ \bibnamefont {Putris}}, \ and\
  \bibinfo {author} {\bibfnamefont {A.}~\bibnamefont {Barry}},\ }\href
  {\doibase 10.1103/PhysRevLett.80.3815} {\bibfield  {journal} {\bibinfo
  {journal} {Phys. Rev. Lett.}\ }\textbf {\bibinfo {volume} {80}},\ \bibinfo
  {pages} {3815} (\bibinfo {year} {1998})}\BibitemShut {NoStop}%
\bibitem [{\citenamefont {Kresse}\ and\ \citenamefont
  {Furthm\"uller}(1996)}]{vasp}%
  \BibitemOpen
  \bibfield  {author} {\bibinfo {author} {\bibfnamefont {G.}~\bibnamefont
  {Kresse}}\ and\ \bibinfo {author} {\bibfnamefont {J.}~\bibnamefont
  {Furthm\"uller}},\ }\href {\doibase
  https://doi.org/10.1016/0927-0256(96)00008-0} {\bibfield  {journal} {\bibinfo
   {journal} {Computational Materials Science}\ }\textbf {\bibinfo {volume}
  {6}},\ \bibinfo {pages} {15 } (\bibinfo {year} {1996})}\BibitemShut {NoStop}%
\bibitem [{\citenamefont {Kresse}\ and\ \citenamefont {Joubert}(1999)}]{paw}%
  \BibitemOpen
  \bibfield  {author} {\bibinfo {author} {\bibfnamefont {G.}~\bibnamefont
  {Kresse}}\ and\ \bibinfo {author} {\bibfnamefont {D.}~\bibnamefont
  {Joubert}},\ }\href {\doibase 10.1103/PhysRevB.59.1758} {\bibfield  {journal}
  {\bibinfo  {journal} {Phys. Rev. B}\ }\textbf {\bibinfo {volume} {59}},\
  \bibinfo {pages} {1758} (\bibinfo {year} {1999})}\BibitemShut {NoStop}%
\bibitem [{\citenamefont {Perdew}\ \emph {et~al.}(1996)\citenamefont {Perdew},
  \citenamefont {Burke},\ and\ \citenamefont {Ernzerhof}}]{GGA}%
  \BibitemOpen
  \bibfield  {author} {\bibinfo {author} {\bibfnamefont {J.~P.}\ \bibnamefont
  {Perdew}}, \bibinfo {author} {\bibfnamefont {K.}~\bibnamefont {Burke}}, \
  and\ \bibinfo {author} {\bibfnamefont {M.}~\bibnamefont {Ernzerhof}},\ }\href
  {\doibase 10.1103/PhysRevLett.77.3865} {\bibfield  {journal} {\bibinfo
  {journal} {Phys. Rev. Lett.}\ }\textbf {\bibinfo {volume} {77}},\ \bibinfo
  {pages} {3865} (\bibinfo {year} {1996})}\BibitemShut {NoStop}%
\bibitem [{\citenamefont {Dudarev}\ \emph {et~al.}(1998)\citenamefont
  {Dudarev}, \citenamefont {Botton}, \citenamefont {Savrasov}, \citenamefont
  {Humphreys},\ and\ \citenamefont {Sutton}}]{Dudarev}%
  \BibitemOpen
  \bibfield  {author} {\bibinfo {author} {\bibfnamefont {S.~L.}\ \bibnamefont
  {Dudarev}}, \bibinfo {author} {\bibfnamefont {G.~A.}\ \bibnamefont {Botton}},
  \bibinfo {author} {\bibfnamefont {S.~Y.}\ \bibnamefont {Savrasov}}, \bibinfo
  {author} {\bibfnamefont {C.~J.}\ \bibnamefont {Humphreys}}, \ and\ \bibinfo
  {author} {\bibfnamefont {A.~P.}\ \bibnamefont {Sutton}},\ }\href {\doibase
  10.1103/PhysRevB.57.1505} {\bibfield  {journal} {\bibinfo  {journal} {Phys.
  Rev. B}\ }\textbf {\bibinfo {volume} {57}},\ \bibinfo {pages} {1505}
  (\bibinfo {year} {1998})}\BibitemShut {NoStop}%
\bibitem [{\citenamefont {Lutfalla}\ \emph {et~al.}(2011)\citenamefont
  {Lutfalla}, \citenamefont {Shapovalov},\ and\ \citenamefont
  {Bell}}]{uvaryref1}%
  \BibitemOpen
  \bibfield  {author} {\bibinfo {author} {\bibfnamefont {S.}~\bibnamefont
  {Lutfalla}}, \bibinfo {author} {\bibfnamefont {V.}~\bibnamefont
  {Shapovalov}}, \ and\ \bibinfo {author} {\bibfnamefont {A.~T.}\ \bibnamefont
  {Bell}},\ }\href {\doibase 10.1021/ct200202g} {\bibfield  {journal} {\bibinfo
   {journal} {Journal of Chemical Theory and Computation}\ }\textbf {\bibinfo
  {volume} {7}},\ \bibinfo {pages} {2218} (\bibinfo {year} {2011})}\BibitemShut
  {NoStop}%
\bibitem [{\citenamefont {Budai}\ \emph {et~al.}(2014)\citenamefont {Budai},
  \citenamefont {Hong}, \citenamefont {Manley}, \citenamefont {Specht},
  \citenamefont {Li}, \citenamefont {Tischler}, \citenamefont {Abernathy},
  \citenamefont {Said}, \citenamefont {Leu}, \citenamefont {Boatner},
  \citenamefont {McQueeney},\ and\ \citenamefont {Delaire}}]{uvaryref2}%
  \BibitemOpen
  \bibfield  {author} {\bibinfo {author} {\bibfnamefont {J.~D.}\ \bibnamefont
  {Budai}}, \bibinfo {author} {\bibfnamefont {J.}~\bibnamefont {Hong}},
  \bibinfo {author} {\bibfnamefont {M.~E.}\ \bibnamefont {Manley}}, \bibinfo
  {author} {\bibfnamefont {E.~D.}\ \bibnamefont {Specht}}, \bibinfo {author}
  {\bibfnamefont {C.~W.}\ \bibnamefont {Li}}, \bibinfo {author} {\bibfnamefont
  {J.~Z.}\ \bibnamefont {Tischler}}, \bibinfo {author} {\bibfnamefont {D.~L.}\
  \bibnamefont {Abernathy}}, \bibinfo {author} {\bibfnamefont {A.~H.}\
  \bibnamefont {Said}}, \bibinfo {author} {\bibfnamefont {B.~M.}\ \bibnamefont
  {Leu}}, \bibinfo {author} {\bibfnamefont {L.~A.}\ \bibnamefont {Boatner}},
  \bibinfo {author} {\bibfnamefont {R.~J.}\ \bibnamefont {McQueeney}}, \ and\
  \bibinfo {author} {\bibfnamefont {O.}~\bibnamefont {Delaire}},\ }\href
  {\doibase 10.1038/nature13865} {\bibfield  {journal} {\bibinfo  {journal}
  {Nature}\ }\textbf {\bibinfo {volume} {515}},\ \bibinfo {pages} {535}
  (\bibinfo {year} {2014})}\BibitemShut {NoStop}%
\bibitem [{\citenamefont {Lu}\ \emph {et~al.}(2019)\citenamefont {Lu},
  \citenamefont {Guo},\ and\ \citenamefont {Robertson}}]{uvaryref3}%
  \BibitemOpen
  \bibfield  {author} {\bibinfo {author} {\bibfnamefont {H.}~\bibnamefont
  {Lu}}, \bibinfo {author} {\bibfnamefont {Y.}~\bibnamefont {Guo}}, \ and\
  \bibinfo {author} {\bibfnamefont {J.}~\bibnamefont {Robertson}},\ }\href
  {\doibase 10.1103/PhysRevMaterials.3.094603} {\bibfield  {journal} {\bibinfo
  {journal} {Phys. Rev. Materials}\ }\textbf {\bibinfo {volume} {3}},\ \bibinfo
  {pages} {094603} (\bibinfo {year} {2019})}\BibitemShut {NoStop}%
\bibitem [{\citenamefont {McWhan}\ \emph {et~al.}(1974)\citenamefont {McWhan},
  \citenamefont {Marezio}, \citenamefont {Remeika},\ and\ \citenamefont
  {Dernier}}]{mcwhan}%
  \BibitemOpen
  \bibfield  {author} {\bibinfo {author} {\bibfnamefont {D.~B.}\ \bibnamefont
  {McWhan}}, \bibinfo {author} {\bibfnamefont {M.}~\bibnamefont {Marezio}},
  \bibinfo {author} {\bibfnamefont {J.~P.}\ \bibnamefont {Remeika}}, \ and\
  \bibinfo {author} {\bibfnamefont {P.~D.}\ \bibnamefont {Dernier}},\ }\href
  {\doibase 10.1103/PhysRevB.10.490} {\bibfield  {journal} {\bibinfo  {journal}
  {Phys. Rev. B}\ }\textbf {\bibinfo {volume} {10}},\ \bibinfo {pages} {490}
  (\bibinfo {year} {1974})}\BibitemShut {NoStop}%
\bibitem [{foo()}]{footnote-SI}%
  \BibitemOpen
  \href@noop {} {}\bibinfo {note} {For additional details about optimized
  crystal structures, Wannier functions spread, Hybrid-functional based density
  of states plot and magnetic configurations and their corresponding energies
  for 25\% doped case, please see the Supplementary Information.}\BibitemShut
  {Stop}%
\bibitem [{\citenamefont {Mostofi}\ \emph {et~al.}(2008)\citenamefont
  {Mostofi}, \citenamefont {Yates}, \citenamefont {Lee}, \citenamefont {Souza},
  \citenamefont {Vanderbilt},\ and\ \citenamefont {Marzari}}]{wannier1}%
  \BibitemOpen
  \bibfield  {author} {\bibinfo {author} {\bibfnamefont {A.~A.}\ \bibnamefont
  {Mostofi}}, \bibinfo {author} {\bibfnamefont {J.~R.}\ \bibnamefont {Yates}},
  \bibinfo {author} {\bibfnamefont {Y.-S.}\ \bibnamefont {Lee}}, \bibinfo
  {author} {\bibfnamefont {I.}~\bibnamefont {Souza}}, \bibinfo {author}
  {\bibfnamefont {D.}~\bibnamefont {Vanderbilt}}, \ and\ \bibinfo {author}
  {\bibfnamefont {N.}~\bibnamefont {Marzari}},\ }\href {\doibase
  https://doi.org/10.1016/j.cpc.2007.11.016} {\bibfield  {journal} {\bibinfo
  {journal} {Computer Physics Communications}\ }\textbf {\bibinfo {volume}
  {178}},\ \bibinfo {pages} {685 } (\bibinfo {year} {2008})}\BibitemShut
  {NoStop}%
\bibitem [{\citenamefont {Franchini}\ \emph {et~al.}(2012)\citenamefont
  {Franchini}, \citenamefont {Kov{\'{a}}{\v{c}}ik}, \citenamefont {Marsman},
  \citenamefont {Murthy}, \citenamefont {He}, \citenamefont {Ederer},\ and\
  \citenamefont {Kresse}}]{wannier2}%
  \BibitemOpen
  \bibfield  {author} {\bibinfo {author} {\bibfnamefont {C.}~\bibnamefont
  {Franchini}}, \bibinfo {author} {\bibfnamefont {R.}~\bibnamefont
  {Kov{\'{a}}{\v{c}}ik}}, \bibinfo {author} {\bibfnamefont {M.}~\bibnamefont
  {Marsman}}, \bibinfo {author} {\bibfnamefont {S.~S.}\ \bibnamefont {Murthy}},
  \bibinfo {author} {\bibfnamefont {J.}~\bibnamefont {He}}, \bibinfo {author}
  {\bibfnamefont {C.}~\bibnamefont {Ederer}}, \ and\ \bibinfo {author}
  {\bibfnamefont {G.}~\bibnamefont {Kresse}},\ }\href {\doibase
  10.1088/0953-8984/24/23/235602} {\bibfield  {journal} {\bibinfo  {journal}
  {Journal of Physics: Condensed Matter}\ }\textbf {\bibinfo {volume} {24}},\
  \bibinfo {pages} {235602} (\bibinfo {year} {2012})}\BibitemShut {NoStop}%
\bibitem [{\citenamefont {Quan}\ \emph {et~al.}(2012)\citenamefont {Quan},
  \citenamefont {Pardo},\ and\ \citenamefont {Pickett}}]{oxd_state}%
  \BibitemOpen
  \bibfield  {author} {\bibinfo {author} {\bibfnamefont {Y.}~\bibnamefont
  {Quan}}, \bibinfo {author} {\bibfnamefont {V.}~\bibnamefont {Pardo}}, \ and\
  \bibinfo {author} {\bibfnamefont {W.~E.}\ \bibnamefont {Pickett}},\ }\href
  {\doibase 10.1103/PhysRevLett.109.216401} {\bibfield  {journal} {\bibinfo
  {journal} {Phys. Rev. Lett.}\ }\textbf {\bibinfo {volume} {109}},\ \bibinfo
  {pages} {216401} (\bibinfo {year} {2012})}\BibitemShut {NoStop}%
\bibitem [{\citenamefont {Mahadevan}\ \emph {et~al.}(2001)\citenamefont
  {Mahadevan}, \citenamefont {Terakura},\ and\ \citenamefont {Sarma}}]{co_prl}%
  \BibitemOpen
  \bibfield  {author} {\bibinfo {author} {\bibfnamefont {P.}~\bibnamefont
  {Mahadevan}}, \bibinfo {author} {\bibfnamefont {K.}~\bibnamefont {Terakura}},
  \ and\ \bibinfo {author} {\bibfnamefont {D.~D.}\ \bibnamefont {Sarma}},\
  }\href {\doibase 10.1103/PhysRevLett.87.066404} {\bibfield  {journal}
  {\bibinfo  {journal} {Phys. Rev. Lett.}\ }\textbf {\bibinfo {volume} {87}},\
  \bibinfo {pages} {066404} (\bibinfo {year} {2001})}\BibitemShut {NoStop}%
\bibitem [{\citenamefont {Luo}\ \emph {et~al.}(2007)\citenamefont {Luo},
  \citenamefont {Franceschetti}, \citenamefont {Varela}, \citenamefont {Tao},
  \citenamefont {Pennycook},\ and\ \citenamefont {Pantelides}}]{oxd_state2}%
  \BibitemOpen
  \bibfield  {author} {\bibinfo {author} {\bibfnamefont {W.}~\bibnamefont
  {Luo}}, \bibinfo {author} {\bibfnamefont {A.}~\bibnamefont {Franceschetti}},
  \bibinfo {author} {\bibfnamefont {M.}~\bibnamefont {Varela}}, \bibinfo
  {author} {\bibfnamefont {J.}~\bibnamefont {Tao}}, \bibinfo {author}
  {\bibfnamefont {S.~J.}\ \bibnamefont {Pennycook}}, \ and\ \bibinfo {author}
  {\bibfnamefont {S.~T.}\ \bibnamefont {Pantelides}},\ }\href {\doibase
  10.1103/PhysRevLett.99.036402} {\bibfield  {journal} {\bibinfo  {journal}
  {Phys. Rev. Lett.}\ }\textbf {\bibinfo {volume} {99}},\ \bibinfo {pages}
  {036402} (\bibinfo {year} {2007})}\BibitemShut {NoStop}%
\bibitem [{\citenamefont {Raebiger}\ \emph {et~al.}(2008)\citenamefont
  {Raebiger}, \citenamefont {Lany},\ and\ \citenamefont {Zunger}}]{oxd_state3}%
  \BibitemOpen
  \bibfield  {author} {\bibinfo {author} {\bibfnamefont {H.}~\bibnamefont
  {Raebiger}}, \bibinfo {author} {\bibfnamefont {S.}~\bibnamefont {Lany}}, \
  and\ \bibinfo {author} {\bibfnamefont {A.}~\bibnamefont {Zunger}},\ }\href
  {\doibase 10.1038/nature07009} {\bibfield  {journal} {\bibinfo  {journal}
  {Nature}\ }\textbf {\bibinfo {volume} {453}},\ \bibinfo {pages} {763}
  (\bibinfo {year} {2008})}\BibitemShut {NoStop}%
\bibitem [{\citenamefont {Grau-Crespo}\ \emph {et~al.}(2012)\citenamefont
  {Grau-Crespo}, \citenamefont {Wang},\ and\ \citenamefont
  {Schwingenschl\"ogl}}]{hse-alpha1}%
  \BibitemOpen
  \bibfield  {author} {\bibinfo {author} {\bibfnamefont {R.}~\bibnamefont
  {Grau-Crespo}}, \bibinfo {author} {\bibfnamefont {H.}~\bibnamefont {Wang}}, \
  and\ \bibinfo {author} {\bibfnamefont {U.}~\bibnamefont
  {Schwingenschl\"ogl}},\ }\href {\doibase 10.1103/PhysRevB.86.081101}
  {\bibfield  {journal} {\bibinfo  {journal} {Phys. Rev. B}\ }\textbf {\bibinfo
  {volume} {86}},\ \bibinfo {pages} {081101} (\bibinfo {year}
  {2012})}\BibitemShut {NoStop}%
\bibitem [{\citenamefont {Franchini}(2014)}]{hse-alpha2}%
  \BibitemOpen
  \bibfield  {author} {\bibinfo {author} {\bibfnamefont {C.}~\bibnamefont
  {Franchini}},\ }\href {\doibase 10.1088/0953-8984/26/25/253202} {\bibfield
  {journal} {\bibinfo  {journal} {Journal of Physics: Condensed Matter}\
  }\textbf {\bibinfo {volume} {26}},\ \bibinfo {pages} {253202} (\bibinfo
  {year} {2014})}\BibitemShut {NoStop}%
\bibitem [{\citenamefont {Williams}\ \emph {et~al.}(2009)\citenamefont
  {Williams}, \citenamefont {Butler}, \citenamefont {Mewes}, \citenamefont
  {Sims}, \citenamefont {Chshiev},\ and\ \citenamefont {Sarker}}]{william6}%
  \BibitemOpen
  \bibfield  {author} {\bibinfo {author} {\bibfnamefont {M.~E.}\ \bibnamefont
  {Williams}}, \bibinfo {author} {\bibfnamefont {W.~H.}\ \bibnamefont
  {Butler}}, \bibinfo {author} {\bibfnamefont {C.~K.}\ \bibnamefont {Mewes}},
  \bibinfo {author} {\bibfnamefont {H.}~\bibnamefont {Sims}}, \bibinfo {author}
  {\bibfnamefont {M.}~\bibnamefont {Chshiev}}, \ and\ \bibinfo {author}
  {\bibfnamefont {S.~K.}\ \bibnamefont {Sarker}},\ }\href {\doibase
  10.1063/1.3072033} {\bibfield  {journal} {\bibinfo  {journal} {Journal of
  Applied Physics}\ }\textbf {\bibinfo {volume} {105}},\ \bibinfo {pages}
  {07E510} (\bibinfo {year} {2009})}\BibitemShut {NoStop}%
\bibitem [{\citenamefont {Williams}\ \emph {et~al.}(2012)\citenamefont
  {Williams}, \citenamefont {Sims}, \citenamefont {Mazumdar},\ and\
  \citenamefont {Butler}}]{william7}%
  \BibitemOpen
  \bibfield  {author} {\bibinfo {author} {\bibfnamefont {M.~E.}\ \bibnamefont
  {Williams}}, \bibinfo {author} {\bibfnamefont {H.}~\bibnamefont {Sims}},
  \bibinfo {author} {\bibfnamefont {D.}~\bibnamefont {Mazumdar}}, \ and\
  \bibinfo {author} {\bibfnamefont {W.~H.}\ \bibnamefont {Butler}},\ }\href
  {\doibase 10.1103/PhysRevB.86.235124} {\bibfield  {journal} {\bibinfo
  {journal} {Phys. Rev. B}\ }\textbf {\bibinfo {volume} {86}},\ \bibinfo
  {pages} {235124} (\bibinfo {year} {2012})}\BibitemShut {NoStop}%
\bibitem [{\citenamefont {Mustonen}\ \emph {et~al.}(2016)\citenamefont
  {Mustonen}, \citenamefont {Vasala}, \citenamefont {Chou}, \citenamefont
  {Chen},\ and\ \citenamefont {Karppinen}}]{musto}%
  \BibitemOpen
  \bibfield  {author} {\bibinfo {author} {\bibfnamefont {O.}~\bibnamefont
  {Mustonen}}, \bibinfo {author} {\bibfnamefont {S.}~\bibnamefont {Vasala}},
  \bibinfo {author} {\bibfnamefont {T.-L.}\ \bibnamefont {Chou}}, \bibinfo
  {author} {\bibfnamefont {J.-M.}\ \bibnamefont {Chen}}, \ and\ \bibinfo
  {author} {\bibfnamefont {M.}~\bibnamefont {Karppinen}},\ }\href {\doibase
  10.1103/PhysRevB.93.014405} {\bibfield  {journal} {\bibinfo  {journal} {Phys.
  Rev. B}\ }\textbf {\bibinfo {volume} {93}},\ \bibinfo {pages} {014405}
  (\bibinfo {year} {2016})}\BibitemShut {NoStop}%
\end{thebibliography}%
 
\end{document}